\documentclass[conference,onecolumn]{IEEEtran}

\usepackage[utf8]{inputenc}
\usepackage{amsmath}
\usepackage{amsfonts}
\usepackage{amsthm}

\usepackage{graphicx}
\graphicspath{{Figures/}}
\usepackage{epstopdf}

\usepackage{url}
\usepackage{comment}

\usepackage{todonotes} 
\usepackage{soul} 

\usepackage{xspace}

\usepackage{algorithm}  
\usepackage{algorithmic} 
\usepackage[short,nocomma]{optidef}

\usepackage{cite}

\usepackage{array}
\usepackage{ragged2e}
\newcolumntype{P}[1]{>{\RaggedRight\hspace{0pt}}p{#1}}

\usepackage{multicol}
\usepackage{multirow}

\usepackage{wrapfig}

\ifCLASSOPTIONcompsoc
\usepackage[caption=false, font=normalsize, labelfont=sf, textfont=sf]{subfig}
\else
\usepackage[caption=false, font=footnotesize]{subfig}
\fi

\usepackage{color}
\usepackage{xcolor}

\usepackage{tikz}
\usetikzlibrary{arrows,shapes,positioning,shadows,trees,mindmap}
\usepackage[edges]{forest}
\usetikzlibrary{arrows.meta}
\colorlet{linecol}{black!75}
\usepackage{xkcdcolors} 

\usepackage{tikz}
\usetikzlibrary{backgrounds}
\usetikzlibrary{arrows,shapes}
\usetikzlibrary{tikzmark}
\usetikzlibrary{calc}
\newcommand{\highlight}[2]{\colorbox{#1!17}{$\displaystyle #2$}}

\usepackage{colortbl}

\newcommand{\eg}{{\it e.g.,}\xspace}
\newcommand{\viz}{{\it viz.,}\xspace}

\newcommand{\ie}{{\it i.e.,}\xspace}
\newcommand{\etc}{{\it etc.}}
\newcommand{\ci}{{\it (i) }}
\newcommand{\cii}{{\it (ii) }}
\newcommand{\ciii}{{\it (iii) }}
\newcommand{\civ}{{\it (iv) }}

\newcommand{\ca}{{\it (a) }}
\newcommand{\cb}{{\it (b) }}
\newcommand{\cc}{{\it (c) }}
\newcommand{\cd}{{\it (d) }}

\DeclareMathAlphabet{\mathpzc}{OT1}{pzc}{m}{it}

\newcommand{\checkactparam}[3]{\textsf{CheckAct}$($#1$,$ #2$,$ #3$)$}
\newcommand{\checkact}{\textsf{CheckAct}()\xspace}

\newcommand{\nw}{\xspace{NW}\xspace}
\newcommand{\sw}{\xspace{SW}\xspace}
\newcommand{\pnametee}{SCATE\xspace}

\begin{document}
%
\title{\Huge \normalfont You Can't Always Check What You Wanted: \\[0.10cm]
{\huge \normalfont  Selective Checking and Trusted Execution to Prevent \\\vspace*{-0.25\baselineskip} False Actuations in Cyber-Physical Systems}}

\date{}





\author{\IEEEauthorblockN{Monowar Hasan}
	\IEEEauthorblockA{School of Electrical Engineering \& Computer Science \\
		Washington State University, Pullman, WA, USA\\
		Email: monowar.hasan@wsu.edu}
	\and
	\IEEEauthorblockN{Sibin Mohan}
	\IEEEauthorblockA{Department of Computer Science \\
		 The George Washington University, Washington, DC, USA\\
		Email: sibin.mohan@gwu.edu}}

\maketitle

\thispagestyle{plain}
\pagestyle{plain}

\begin{abstract}
Cyber-physical systems (CPS) are vulnerable to attacks targeting outgoing actuation commands that modify their physical behaviors.
%
The limited resources in such systems, coupled with their stringent timing constraints, often prevents the checking of every outgoing command. 
We present a ``selective checking'' mechanism that uses game-theoretic modeling to identify the right subset of commands to be checked in order to deter an adversary.
This mechanism is coupled with a ``delay-aware'' trusted execution environment (TEE) to ensure that only verified actuation commands are ever sent to the physical system, thus maintaining their safety and integrity. 
The \textit{selective checking and trusted execution (\pnametee)} framework is implemented on an off-the-shelf ARM platform running standard embedded Linux. 
We demonstrate the effectiveness of \pnametee using four realistic cyber-physical systems (a ground rover, a flight controller, a robotic arm and an automated syringe pump) and study design trade-offs.

Not only does \pnametee provide a high level of security and high performance,
it also suffers from significantly lower overheads ($30.48\%$--$47.32\%$ less) in the process. 
In fact, \pnametee can work with more systems 
without negatively affecting the safety of the system.
Considering that most CPS do not have any such checking mechanisms, and \pnametee is \textit{guaranteed} to meet all the timing requirements (\ie ensure the safety/integrity of the system), our methods can significantly improve the security (and, hence, safety) of the system.
\end{abstract}

\section{Introduction} 
\label{sec:intro}

A large number of critical systems that are in operation today (\eg autonomous cars, avionics, drones, power grids, industrial control systems, medical devices, automobiles, space vehicles, \etc) can be classified as ``cyber-physical systems'' (CPS)\footnote{The term cyber-physical systems refers to the tight conjoining of and coordination between computational and physical resources.}.
Such systems often have limited resources (processor, memory, battery life, \etc) and must also meet stringent \textit{timing} constraints. 
For instance, an industrial robot on a manufacturing line, must carry out its operation (\eg placing an object on a conveyor) in $50$--$100$~ms~\cite{castelli2019development}. 
%
Failure to do so, could disrupt the entire manufacturing operation and even put the safety of the plant and human operators at risk!

Modern CPS with such ``real-time'' requirements are increasingly becoming targets for cyber-attacks.
Traditional safety and fault-tolerance mechanisms used in real-time CPS were designed to counter random or accidental faults and failures and cannot deal with intentional cyber attacks orchestrated by an intelligent and capable adversary.
Further, \ci the drive towards the use of standardized protocols and off-the-shelf components (for interoperability, reduced infrastructure and maintenance costs) and 
\cii the emergence of smart and connected systems (\eg real-time Internet-of-things~\cite{mhasan_rtiot_sensors19}, smart grids~\cite{otuoze2018smart}, smart manufacturing~\cite{zheng2018smart}, smart transportation~\cite{azgomi2018brief}, \etc) are increasing the attack surfaces available to adversaries. 
The traditional approaches of air-gapping such systems~\cite{securecore,mhasan_resecure_iot} or using proprietary protocols and hardware~\cite{mohan_s3a} have been found wanting in the face of recent high-profile attacks:
\eg Stuxnet~\cite{stuxnet}, attack demonstrations by researchers on medical devices~\cite{security_medical} and automobiles~\cite{checkoway2011comprehensive}, denial-of-service (DoS) attacks mounted from IoT devices~\cite{ddos_iot_camera}, among others. 

A common thread among all of the aforementioned attacks, especially ones that threaten the physical safety of the system, is the \textit{falsification of actuation commands} --- \ie commands that control the state of the physical system are either modified or replaced while in transit to the physical component.
Note that CPS are comprised of a tight interplay between computation, control and communication. 
At its core, a cyber-physical ``plant'' consists of actuators and sensors that, respectively, monitor and control the physical properties of the system. 
Due to the tight coupling between actuators and physical plants, such systems are often vulnerable to unexpected situations (\eg malicious actions) that were not considered during the design/development phases~\cite{simmon2013vision}. 
Hence, sending false/spoofed commands to the actuators can disrupt the normal operation of the physical plant and jeopardize its safety.
In the manufacturing robot example mentioned earlier, the attacker could modify the command that changes the angle of rotation of the robot arm, thus causing it to completely miss the conveyor and, potentially, crash the entire system.

We intend to \textit{check actuation commands before they can affect the state of the physical system}.
If we find that the command can have negative effects, \ie it compromises the safety and/or integrity of the CPS, then we prevent it from being sent out.\footnote{Note that the actions taken when we detect a problematic actuation command is orthogonal to the work presented here and depends on the specific CPS. 
%
%
%
We briefly discuss some of these strategies in Sec.~\ref{sec:discussion}.}
To prevent tampering, we \textit{implement the checking mechanism in a trusted execution environment (TEE)} that is available on modern commodity processors, \viz the ARM TrustZone~\cite{trustzone_survey}\footnote{This does not preclude the use of other TEEs, \eg Intel SGX~\cite{intel_sgx}.}.
In an ideal scenario, every outgoing actuation command should be checked. 
A serious hurdle that prevents such a strategy is that, as mentioned earlier, CPS have stringent timing requirements --- very often the actuation command, once sent out, \textit{must be received by the physical system in a short, fixed, amount of time}.
This limits the amount of time delays that can be introduced during the checking process. 
In addition, the control software has its own timing constraints, \eg it must complete execution before a certain ``deadline''; failure to do so can also cause instability in the system. 
Hence, we \textit{cannot check (and thus, delay) every command} since each check compounds the delays faced by the system. 
In addition, as seen in Sec.~\ref{sec:req_int_check}, the use of the TEE-based checking mechanism introduces additional delays due to context-switch overheads, further (negatively) affecting the deadlines. 

Hence, there is a need to carefully consider \textit{how many} and \textit{which} actuation commands are to be checked --- to ensure that 
\ca the system safety/timing requirements are met and 
\cb also deter attackers. 
Picking a fixed subset of actuation commands is not helpful since an adversary can circumvent the checks by targeting the ``unchecked'' commands. 
To this end, \textit{we develop a mechanism to validate a random subset of commands, varying at run-time} that significantly increases the difficulty for would-be attackers. 
We use a \textit{game-theoretic formulation} of a two-player normal-form game~\cite{harsanyi1972generalized, conitzer2006computing, paruchuri2007efficient} for the ``selective checking'' of the actuation commands (Sec.~\ref{sec:overload_control}). 
The combined framework is referred to as the \textit{selective checking and trusted environment (\pnametee)} system.

\noindent \textbf{Contributions of this paper:}
\begin{itemize}
    \item we present a framework, \pnametee, that protects cyber-physical systems from attacks that falsify actuation commands. [Sec.~\ref{sec:prob_overview}]
\item we use a combination of game-theoretic analysis and a trusted execution environment to deter attackers, significantly reduce checking overheads and still guarantee the safety and integrity of the CPS. [Sec.~\ref{sec:overload_control}] 
\end{itemize}
We implemented \pnametee [Sec.~\ref{sec:eval}] on a commercially available ARM Cortex A53 platform~\cite{rpi3} and commodity TEE (TrustZone~\cite{trustzone_survey}) running embedded Linux.
Our system and techniques were evaluated using \textit{four realistic, standardized, cyber-physical systems} [Sec.~\ref{sec:cs_experiments}]:
\ca an autonomous ground rover, 
\cb a flight controller, 
\cc a robotic arm (typically found in manufacturing systems) and
\cd a syringe pump used in medical devices. 
We also carried out a broader design space exploration [Sec.~\ref{sec:syn_experiments}] using simulated workloads and also analyzed the trade-offs between security and timing/safety properties. 
Not only does \pnametee deter attacks, it is able to do so with significantly fewer overheads and also \textit{guarantee} that it will not compromise the system safety and timing properties. 
Our open-sourced implementation is available in a public repository~\cite{rt_act_sec_game_repo}.

\section{Motivation, Overview and Background} \label{sec:mot_ov_bg}




\subsection{The Requirement for Checking Actuation Commands} \label{sec:motivation}




Real-time CPS consists of \textit{cyber}
components and \textit{physical} components. The cyber units perform the computations for estimating the system state (\eg current location of an unmanned vehicle and the intended direction of its movement) and generation of appropriate control signals for the actuators. The physical components include the entities that are \ci closer to the physical system such as sensors that take measurements and \cii the actuators that move the physical system. 

\begin{figure}
	\centering
	\includegraphics[scale=0.28]{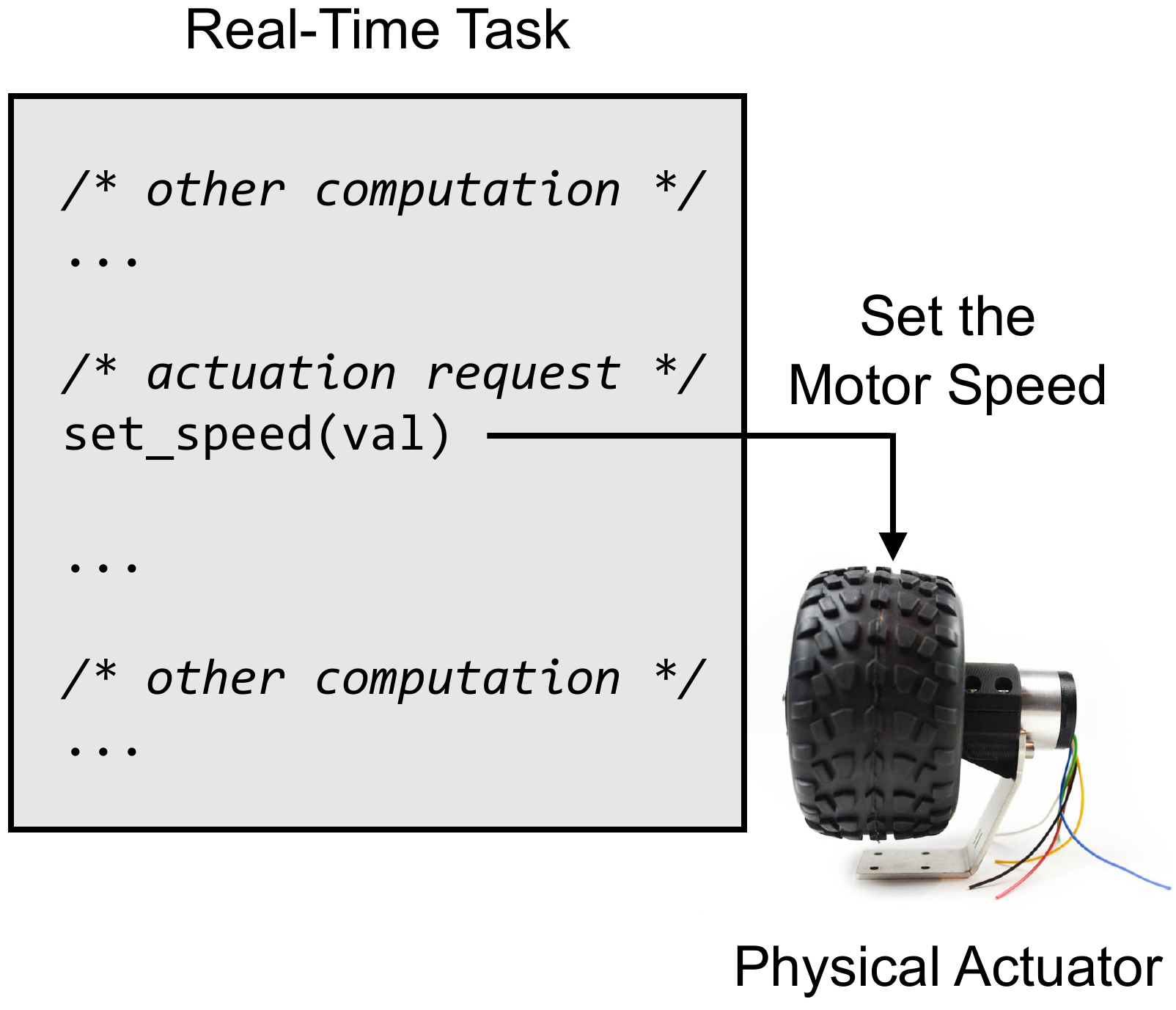}
	\caption{Vanilla (non-secure) execution: does not check actuation commands.} 
	\label{fig:steps_vanilla}
\end{figure}

In a non-secure system, when a task generates an actuation command, it is directly issued to the physical actuators. Figure~\ref{fig:steps_vanilla} illustrates this: when a controller task generates an actuation command (say, to set the speed of the motor), the speed is changed without checking whether the provided speed value is legitimate or not. Without explicit control and verification over the actuation process, it is possible to send arbitrary signals to the actuators and an adversary can drive the system in undesirable ways. For instance, consider ground rovers that can be used in multiple cyber-physical applications such as remote surveillance, agriculture and manufacturing~\cite{guo2018roboads}. For demonstration purposes, we use a COTS-based ground rover 
running an embedded variant of Linux on an ARM Cortex-A53 platform (Raspberry Pi~\cite{rpi3}). The rover is equipped with two optical encoders that are connected to the motors (\ie actuator in this setup); it can turn left by switching off the right encoder and vice-versa. 

\begin{figure} 
	\centering
	\subfloat[\label{fig:rover_attack_illust_a}]{%
		\includegraphics[scale=0.053]{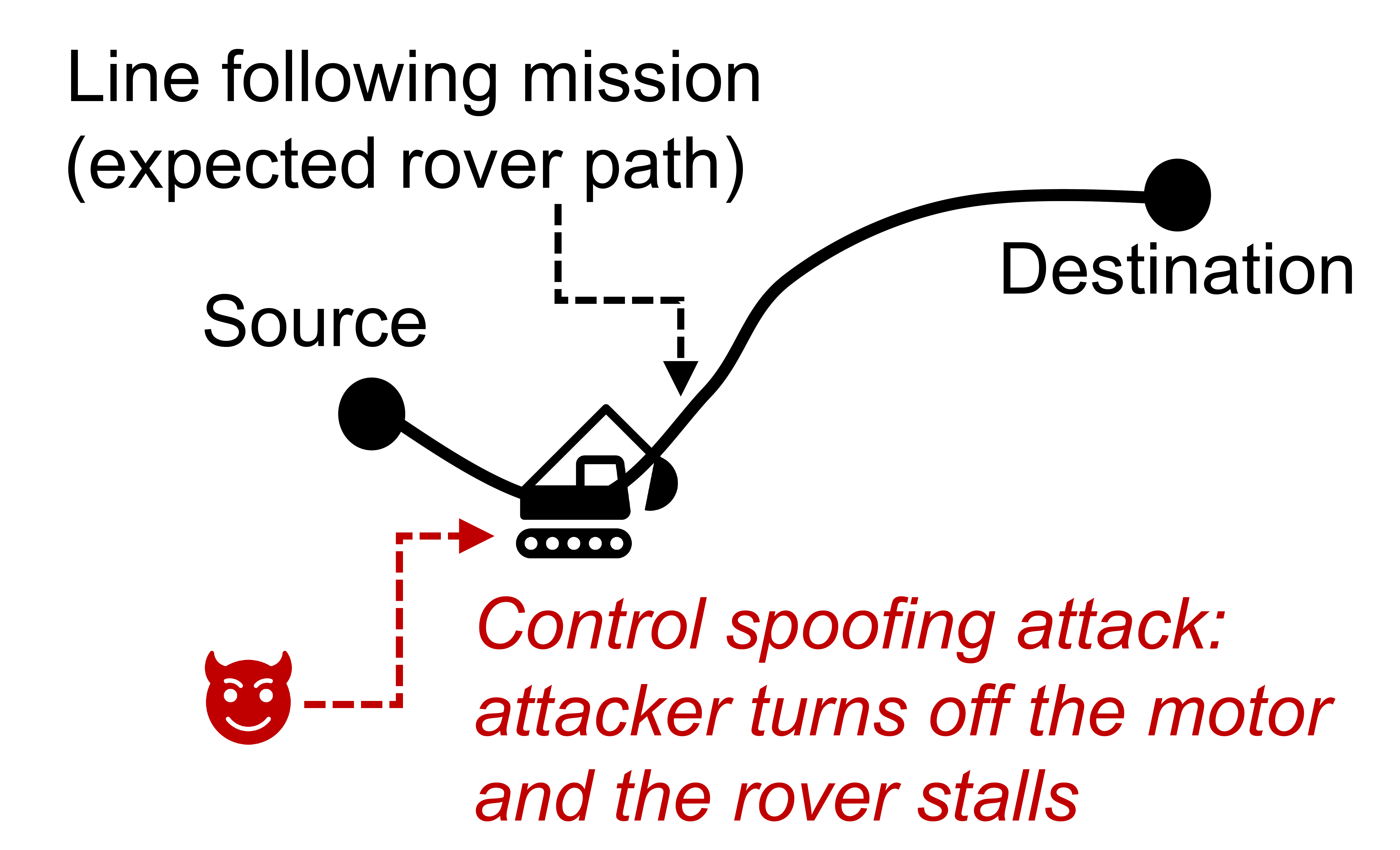}}
	\hspace*{0.2em}
	\subfloat[\label{fig:rover_attack_illust_b}]{%
		\includegraphics[scale=0.36]{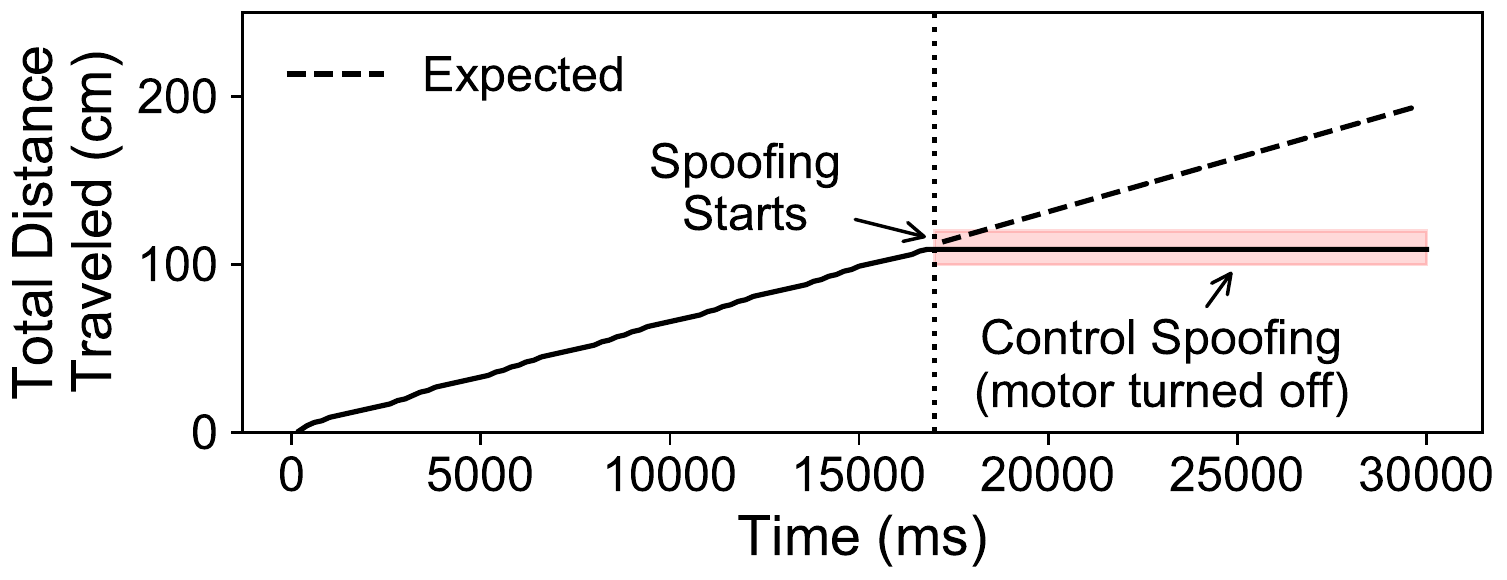}}
	\caption{Illustration of control spoofing attacks on a rover platform: (a) high-level schematic of our experiment setup where a rover performs a line following mission and an adversary triggers malicious code that turns off one of the rover motors; (b) readings from rover motor encoders under control spoofing attack. Without proper control checking, an adversary can inject erroneous signals (shaded region in Fig.~\ref{fig:rover_attack_illust_b}) and deviate the rover from its expected behavior (dashed line). In this setup the rover stalled in the middle of its mission due to control spoofing attack as shown by the constant readings from wheel motors).}
	\label{fig:rover_attack_illust}
\end{figure}

As depicted in Fig.~\ref{fig:rover_attack_illust_a}, we carried-out a line-following mission where the rover steered from an initial location to a target location by following a line. A controller task runs the standard, pre-packaged, proportional–integral–derivative (PID) closed-loop control~\cite{gpg2_lf}. 
A $5$-byte value is sent to the actuator (via memory-mapped registers) to control the wheel motors via the I2C interface~\cite{i2c}. This aids in the navigation and control of the rover. The x-axis of Fig.~\ref{fig:rover_attack_illust} shows the time and the y-axis is the total distance traveled by the rover (\ie readings from the wheel motors). Since the vendor implementation of the controller does not verify control commands, we were able to inject a logic bomb and send spoofed commands to turn off the motor (at the location marked in the figure). As a result, the rover deviates from its mission. The dashed line (after $t=17$ seconds) shows the expected behavior (\viz without any attack) as a reference (obtained by running a linear regression test from the traces of an uncompromised execution). As the shaded region in the figure shows,  the encoder readings (\ie traversed distance) remain the same and the rover was not following the line after the attack. 

We note that designing and scheduling checking techniques for real-time cyber-physical platforms is often more challenging when compared to the general-purpose systems due to additional timing/safety constraints imposed by such systems (see Table~\ref{tab:eval_plat_overview} for related examples). To the best of our knowledge, there is no existing technique that can be directly retrofitted to solve this problem. We believe there is a practical requirement and intellectual merit for designing a framework that can protect actuators in real-time CPS and hence is the focus of this research.








\subsection{System Model} \label{sec:sys_mod}

We consider a set of priority-driven, periodic real-time tasks, $\Gamma$, running on a multicore CPS platform $\Pi$. The set of tasks $\Gamma_p \subset \Gamma$ running on a given core $\pi_p \in \Pi$ is fixed and given by the designers. Each task $\tau_i$ issues $N_i$ number of actuation requests. We assume that there is a designer-given quality-of-service (QoS) requirement that $N_i^{min} \leq N_i$ actuation requests (among total $N_i$ number of  requests) must be checked (by the trusted entity running inside TEE) for each invocation of a task.  We further assume that each actuation command $a_i^j$ is associated a designer-provided weight $\omega_i^j$ that represents the importance/preference of checking the corresponding command over other. A higher weight implies that the actuation request is more critical and designers want to examine it more often. We provide a formal representation of our task and real-time model in Appendix~\ref{sec:sys_mod_formal}.

\subsection{Adversary Model} \label{sec:adv_mod}

Our assumptions on adversarial capabilities is similar to that considered in prior work~\cite{mahfouzi2019butterfly, mhasan_iotsnp19}. In particular, we assume that an adversary can tamper with the existing control logic to manipulate actuation commands, thus modifying the behavior of a system in undesirable
ways (\ie threaten the safety of the system). We only consider the cases where an adversary's actions results in the modification of actuation commands. Other classes of attacks such as scheduler side-channel attacks~\cite{cy_scheduleak_rtas19,liu2019leaking}, timing anomalies~\cite{mahfouzi2019butterfly,dos_heechal_rtas19} and network-level man-in-the-middle attacks~\cite{lesi2017network,lesi2017security} are not within the scope of this work. However, we do discuss how our approach can be extended to other use-cases and mitigate some of those attacks (Sec.~\ref{sec:discussion}). We do
not make any assumptions as to how an adversary compromises tasks or actuation commands. 
For instance, bad software engineering practices leave vulnerabilities in the systems~\cite{loi2017systematically}. When the system is developed using a multi-vendor model~\cite{sg2} (where various components are manufactured and integrated by different vendors) malicious code may be injected (say by a less-trusted vendor) during deployment. The adversary may also induce end-users to download modified source code and/or remote access Trojans, say by using
social engineering tactics~\cite{securecore_syscal}. We note that embedded real-time CPS have fewer resources and lesser security protections and hence make some of the attacks easier for the adversary~\cite{mahfouzi2019butterfly,kim2018securing,cy_scheduleak_rtas19,choi2018detecting}. We do not consider the adversarial cases that require physical access, \ie the attacker can not physically control/turn off/damage the actuators or the system.

\subsection{Problem Overview and Our Approach} \label{sec:prob_overview}



%

\begin{figure}[!t]
	\centering
	\includegraphics[scale=0.29]{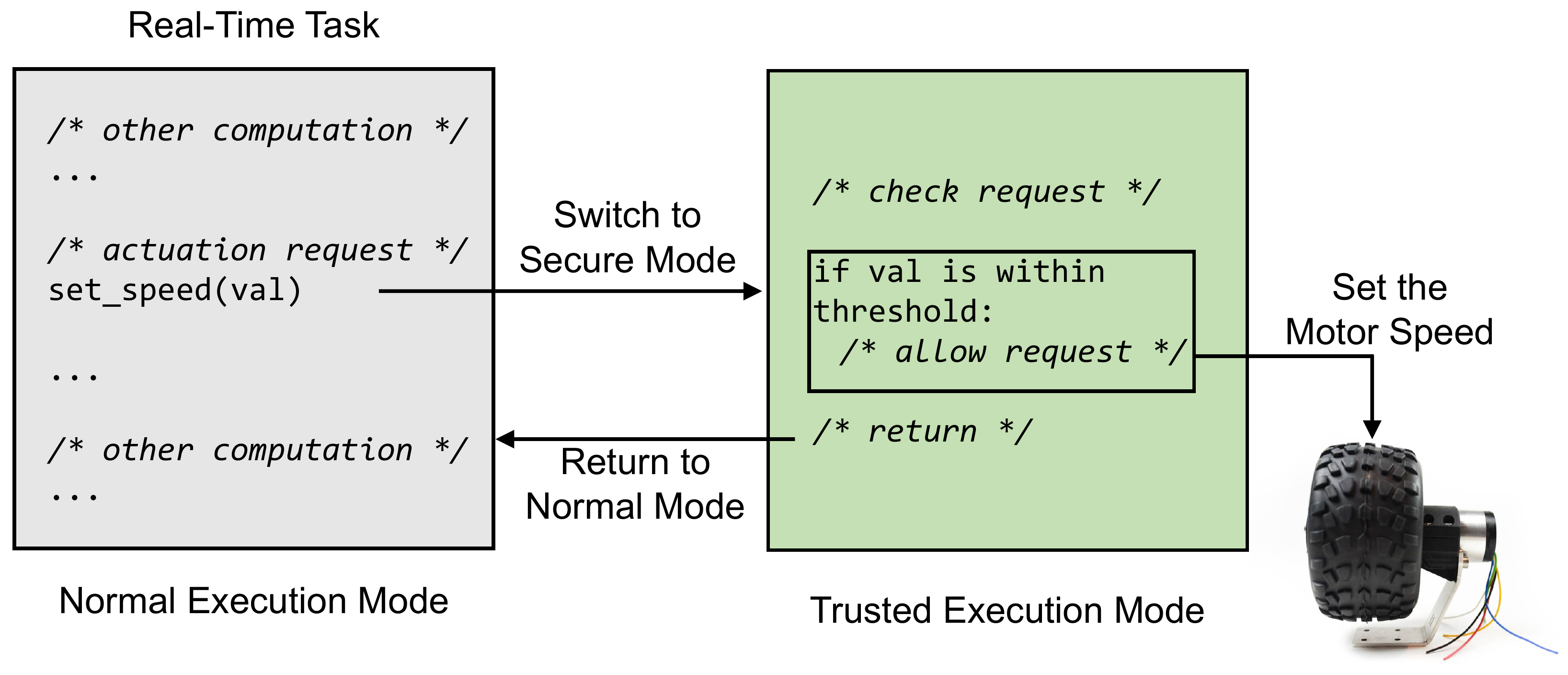}
	\caption{Flow of operation in \pnametee.  When a task generates any actuation command, it will be checked by a trusted entity leveraging TEE technologies (say ARM TrustZone~\cite{trustzone_survey}). In this illustration the speed of the actuators (motors attached to the wheels) is only set if the speed value is within a predefined range.} 
	\label{fig:framework_steps}
\end{figure}

\begin{figure}[!t]
		\centering
	\hspace*{-1em}
	\includegraphics[scale=0.2]{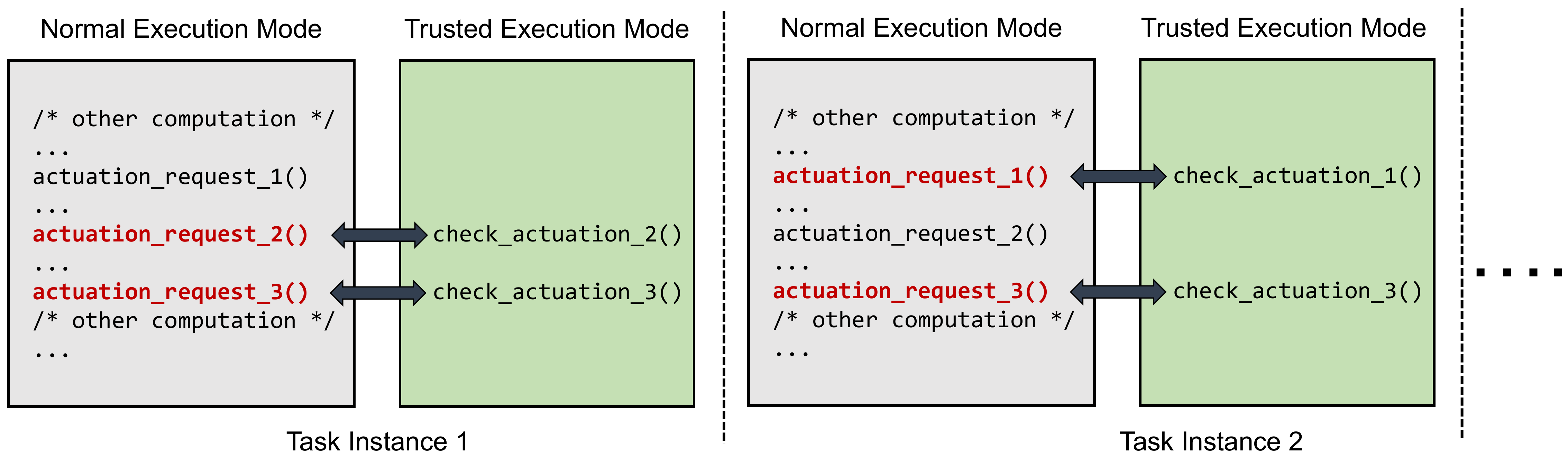}
	\caption{Random selection of actuation commands for checking.} 
	\label{fig:framework_steps_2}
\end{figure}

Actuation commands that are malicious can jeopardize the safety and integrity of cyber-physical applications. In this research we propose techniques to \textit{protect systems from control spoofing attacks by examining actuation commands} \textbf{before} they are issued to physical peripherals. We name our mechanism, \pnametee (selective checking and trusted execution) where
 we consider cyber-physical applications consisting of software tasks that can have two different types of execution sections: \ca regular (potentially untrusted) execution section where normal executions are carried-out and \cb trusted sections where critical information (\ie actuation events) are examined. 
 
 The high-level schematic of \pnametee is depicted in Fig.~\ref{fig:framework_steps}. When a task issues an actuation command, \pnametee transfers the control to the secure mode using the TrustZone secure call (SMC) instructions~\cite{trustzone_survey}. In the secure mode a designer-provided trusted entity checks the actuation commands. In this work we assume that our checking module uses policy rules that defines the mapping of various system states to corresponding legitimate actuation events (refer to Sec.~\ref{sec:act_verif_n_timing} for details). We note that although we use ARM TrustZone as the underlying TEE to demonstrate our ideas, other trusted environments can be used in \pnametee  without loss of generality. 
 
 As we shall see in Sec.~\ref{sec:req_int_check}, the context switch overhead for switching between normal and trusted modes is not negligible.  For example, consider the rover used in our experiments (Sec.~\ref{sec:eval} and Table~\ref{tab:eval_plat_overview}). The execution frequency\footnote{Note: these tasks are periodic in nature.} of the controller task  is $5$ Hz ($200$ ms) and it generates four actuation commands (to set the speeds and direction of attached motors). The controller task must complete execution before its periodic invocation interval ($200$ ms). If we check the speed and direction values of each of the four commands (using TrustZone), the controller task fails to comply with its timing requirements\footnote{This may cause the rover to deviate from its mission, destabilize or even crash.} (since it requires $261$ ms to finish). 
 For such situations (\ie when not all the commands can be verified without missing timing guarantees for all the tasks), \pnametee \textit{selectively checks a subset of the actuation events}. For this, we leverage the tools from \textit{game theory}~\cite{conitzer2006computing,harsanyi1972generalized} and randomly select a subset of commands (for checking) that provides us a trade-off between security and timing guarantees (see Sec.~\ref{sec:overload_control} for details).  Figure~\ref{fig:framework_steps_2} shows an illustrative case where a task generates three actuation requests and we can check at most two request to comply with timing/safety requirements. In this case \pnametee randomly checks two commands each time (\eg commands $[2,3]$ in first instance and commands $[1,3]$ in the second instance). 
 From the earlier rover example, by reducing the number of checks in half\footnote{Section~\ref{sec:eval} presents details of this experimental setup and additional results.} (\eg randomly checking two commands in each task instances instead of all four), the controller task in  \pnametee manages to finish before $200$ ms without significantly degrading security as explained in Sec.~\ref{sec:eval}.

\subsection{Background} \label{sec:background}

We now provide a background on real-time CPS, the TEE technology (ARM TrustZone~\cite{trustzone_survey}) and the game-theoretic modeling~\cite{conitzer2006computing}) used in our work.

\subsubsection{Real-Time Cyber-Physical Systems}

\begin{wrapfigure}{r}{0.30\textwidth}
	\centering
	\includegraphics[scale=0.12]{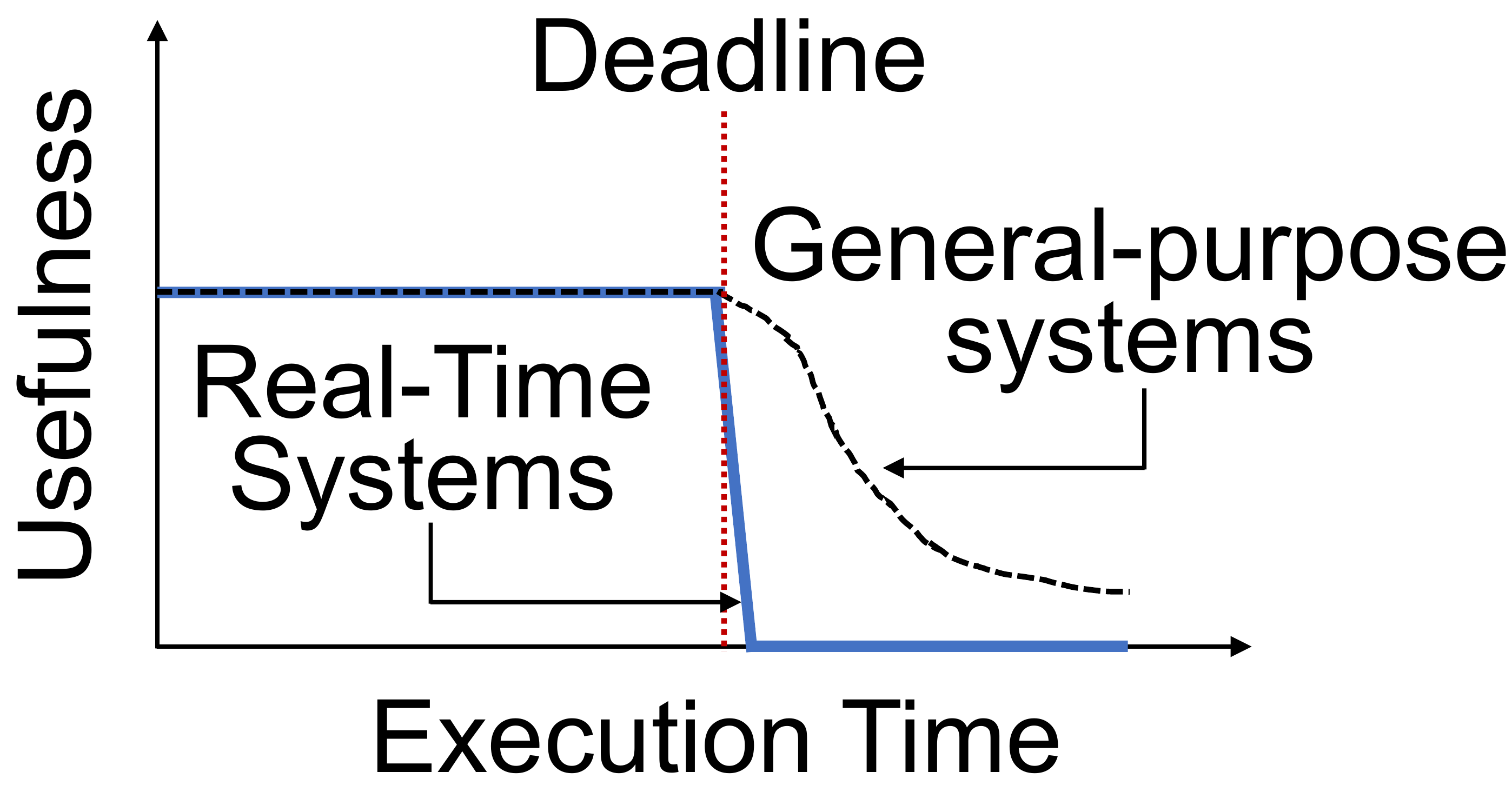}
	\caption{Timeliness requirements of real-time systems.} 
	\label{fig:rt_deadline}
\end{wrapfigure}

Systems with real-time properties need to function correctly, but \textit{within their predefined timing constraints}, often termed as a ``deadlines''. The usefulness of results produced by the system usefulness drops sharply (see Fig.~\ref{fig:rt_deadline}). This is different from general-purpose systems where the usefulness drops in a more gradual manner (\eg a web service may tolerate a few millisecond delays without significantly degrading user experience). For example, consider the CPS use-case of a car's airbag deployment, where the system must react (\ie detect the collision and deploy airbags)  within $60$ ms~\cite{hussain2006vehicle}. If the airbag deployment process fails to response within its timing constraints (\eg $60$ ms), it can have catastrophic results and threaten the safety of the passengers.
Each real-time application in the system (called ``task'') represents a time-critical function such as estimating system state/dynamics or issuing requests to physical peripherals. 
The scheduler in a real-time operating system (RTOS) uses timers and interrupt handlers to enforce timing guarantees (deadlines) at runtime. Some of the common properties and assumptions related to real-time CPS are as follows: 
    \ci implemented as a system of periodic tasks that periodically perform sensing/actuation operations
    \cii memory and processing power often limited and
    \ciii stringent timing and safety requirements.


\subsubsection{ARM TrustZone}

Trusted environments are set of hardware and software-based security extensions where the processors maintain a separated
subsystem in addition to the traditional
OS (also called rich OS) components. TEE technology has been implemented
on commercial hardware such as ARM TrustZone~\cite{trustzone_survey}
and Intel SGX~\cite{intel_sgx}. In this work we consider TrustZone as the building block of our model due to the wide usage of ARM processors in CPS.  We note that although we use the TrustZone functionality for demonstration purposes, our ideas are rather general and can be adapted to other TEE technology without loss of generality.


TrustZone contains two different privilege
blocks: \ci regular (non-secure) execution environment, called ``Normal World'' (\nw) and
\cii a trusted environment, referred to as ``Secure World'' (\sw). The \nw is the untrusted environment that runs a commodity untrusted OS (called rich OS) whereas \sw is a protected computing block that only
runs privileged instructions.  
TrustZone hardware ensures that the resources in the \sw can not be accessed from the \nw.
 These two worlds are bridged via a software module, the \textit{secure monitor}.
The context switch between the \nw and \sw is performed via a \textit{secure monitor call} (SMC).

\subsubsection{Normal-Form Games}

 The overheads for TEE context switch is costly (Sec.~\ref{sec:req_int_check}). \textit{If a task cannot verify all the actuation commands, we propose to select only a subset of commands in each job} for checking. For this, we leverage the tools from game theory~\cite{algorithmic_game_theory} to ensure that the chosen subsets are non-deterministic, at least from the adversary's point of view (see Sec.~\ref{sec:overload_control} for details). In multi-agent systems, if the optimal action for one agent depends on the actions that the other agents take, game theory is used to analyze how an agent should behave in such settings.  In a \textit{normal-form game}~\cite{conitzer2006computing}, every player $j \in \{1, 2, \cdots, J \}$ has a
set of strategies (or actions) $\sigma_j$ and a utility function
$u_j : \sigma_1 \times \sigma_2 \times \cdots \times \sigma_J \rightarrow \mathbb{R}$ that maps every outcome (a vector
consisting of a strategy for every player) to a real number. As we shall see in Sec.~\ref{sec:game_forumlation} we formulate our problem as a \textit{two-player game} (\eg system designer and adversary). 
The output of the game finds the probability distribution over the player's strategies (\ie fraction of time a given strategy is selected in the game) that leads to optimal outcome. While game-theoretic analysis has been used in other modeling problems (\eg patrolling~\cite{paruchuri2007efficient}, network routing~\cite{korilis1997achieving}, transportation systems~\cite{cardinal2005pricing})
as well as general-purpose control systems/CPS~\cite{moothedath2020game, yang2019adaptive, chen2016game, rass2017physical}, they are not real-time aware and do not consider the problem of protecting physical actuators. To the best of our knowledge this is the first work that uses normal-form games in the real-time security context.



\section{Checking Actuation Commands} 
\label{sec:act_verif_n_timing}


\begin{figure} 
	\centering
	\subfloat[\label{fig:state_action_map}]{%
	\includegraphics[scale=0.11]{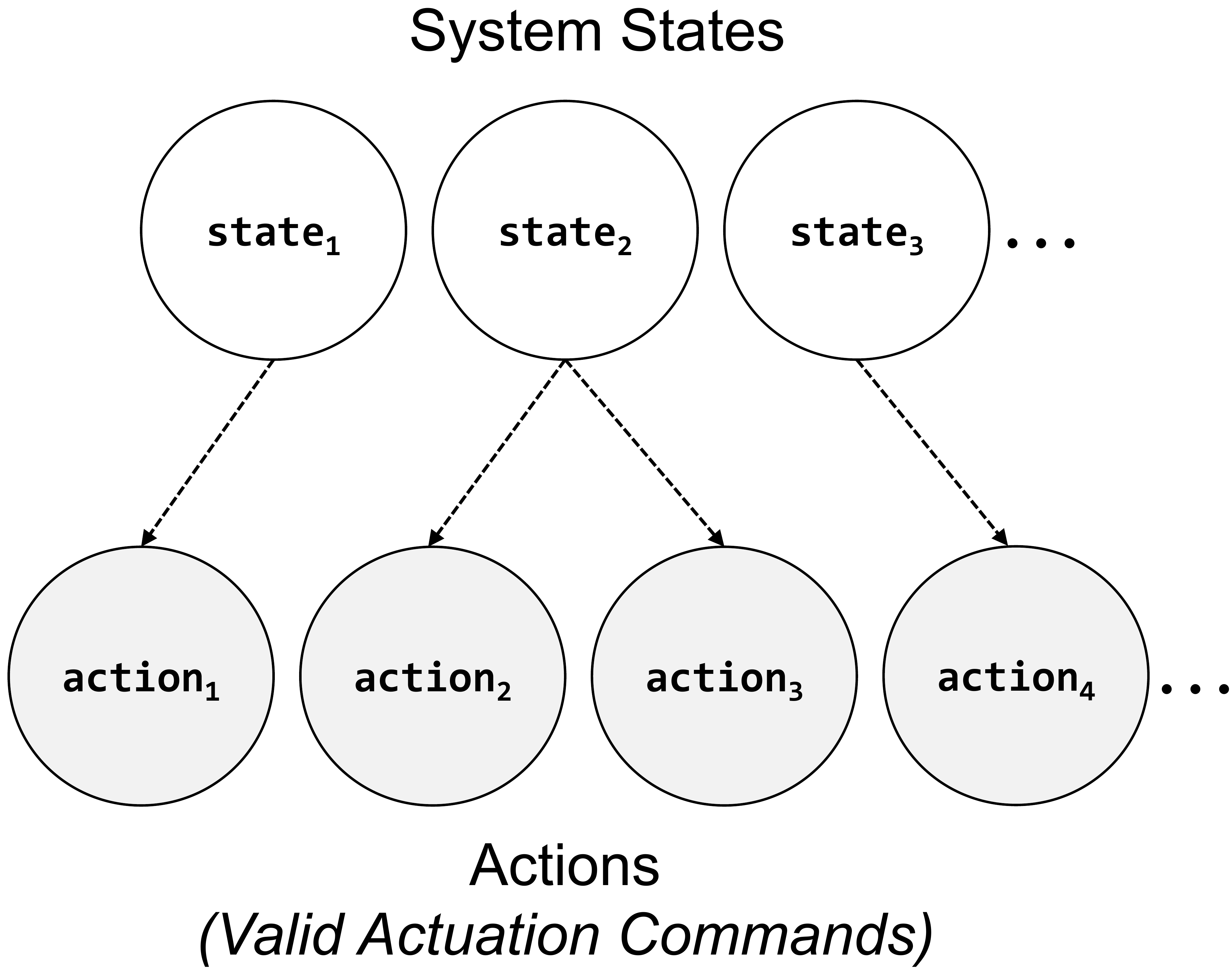}}
 	\hspace*{0.2em}
	\subfloat[\label{fig:state_action_map_rover}]{%
		\includegraphics[scale=0.12]{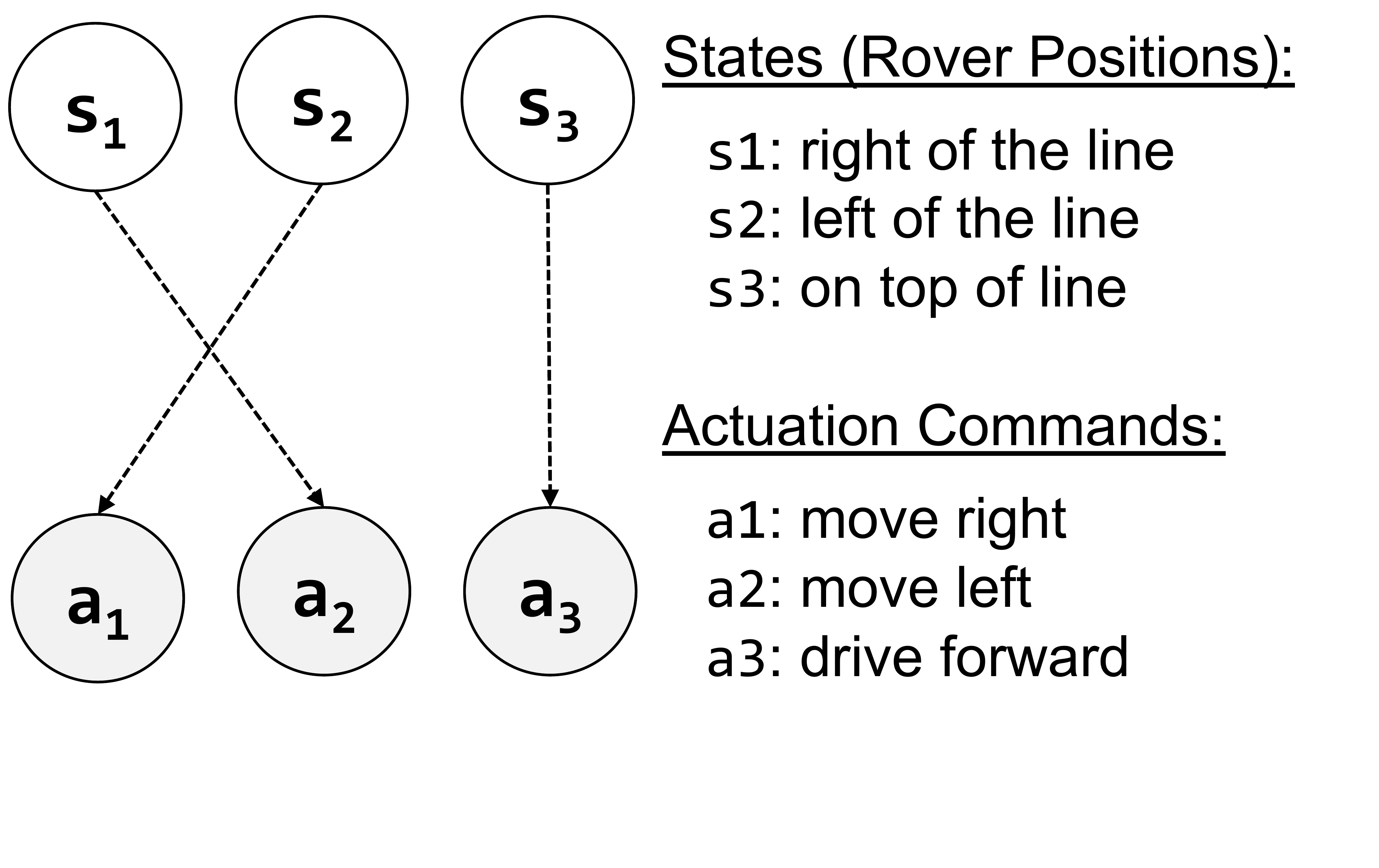}}
	\caption{\ca $State \rightarrow  Action$ mapping used in \pnametee for checking actuation commands. \cb Example states and corresponding valid actuation commands for a line-following rover.}
	\label{fig:state_action}
\end{figure}






In \pnametee, a ``checking module'' executes inside the trusted environment. 
The checking module observes system states and decides whether a given actuation command is legitimate or malicious. 
Recall that when a task issues actuation requests, we transfer control to the secure execution mode using TrustZone SMC instructions. 
In particular, for a given real-time platform we assume that there exists a \checkactparam{$\tau_i$}{$a_i^j$}{$t$} function\footnote{The exact function depends on the specific CPS and application requirements.} that examines a given actuation command $a_i^j$ (where $1\leq j\leq N_i$, $N_i$ is total number of commands the task issues) generated by a task $\tau_i$ at a given time $t$.  
As shown in Fig.~\ref{fig:state_action_map}, the checking module uses policy abstraction rules~\cite{yu2015handling}, \viz $State \rightarrow  Action$ pairs where the $State$ predicate represents a given system state and $Action$ denotes corresponding valid actuation command(s). 
In particular, we assume that when $\tau_i$ executes an actuation command, function \checkactparam{$\tau_i$}{$a_i^j$}{$t$} first observes the system state  $\mathcal{S}(t)$ 
and then decides whether the actuation command $a_i^j$ is valid for the current state $\mathcal{S}(t)$. 

Consider the line-following rover presented in Sec.~\ref{sec:motivation}. 
The directions for the wheels of the rover (\ie forward, left and right; controlled by the attached motors) are the actuation commands.  
At any given point in time, the rover can be in one of three states: $\mathcal{S}=\{ \mathrm{ON\_LINE}, \mathrm{LEFT\_OF\_LINE}, \mathrm{RIGHT\_OF\_LINE}\}$ that denotes whether the rover is on top of the line or shifted left/right of the line, respectively. 
The rover controller task performs the following actuation operations (\ie actions): 
$\mathtt{move\_left()}$/$\mathtt{move\_right()}$ (move the rover left/right, respectively) and $\mathtt{move\_forward()}$ (drive the rover forward). 
The corresponding $State \rightarrow  Action$ mapping for this rover is illustrated in Fig.~\ref{fig:state_action_map_rover}. 
If the rover is on the left side of the line (\ie $State = \mathrm{LEFT\_OF\_LINE} $), the valid command should be $\mathtt{move\_right()}$ (\ie shift the rover back  so that it stays on the line) and if the rover is on top the line (\ie $State = \mathrm{ON\_LINE} $), the controller task should drive it forward (\ie issue $\mathtt{move\_forward()}$ command). 




Table~\ref{tab:rt_act_exp} in the Appendix~\ref{sec:act_chek_sec} summarizes the possible checks for various real-time platforms --- however, this is by no stretch meant to an exhaustive list. We assume that $State \rightarrow  Action$ rules are given by the designers based on system requirements. We note that the ideas presented here are agnostic to the specific checking method and \pnametee is compatible with existing techniques (\eg defining rules at design times~\cite{adepu2017design,mhasan_iotsnp19}, deriving from specifications~\cite{berthier2011specification} and based on statistical analysis~\cite{anomaly_detection_survey}).
%
%
%
%
%
%
In this work we focus on how to selectively examine random subsets of actuation commands by using such designer-provided checking rules. 
%
In Sec.~\ref{sec:cs_experiments} we describe the implementation of \checkact functions for four realistic platforms used in our evaluation.

\subsection{The Requirement for Coarse-Grain Checking} 
\label{sec:req_int_check}

In order to check actuation commands we must ensure that \pnametee should not cause inordinate delays and the timing requirements of real-time tasks are satisfied (\ie they complete execution before their respective deadlines).  We therefore develop design-time tests (see Appendix~\ref{sec:timing_analysis}) that ensure tasks meet their timing requirements (deadlines). Our analysis in Appendix~\ref{sec:timing_analysis} shows that there is an overhead for inspecting the actuation commands using TEEs and a task may miss its timing/safety requirements as a result.  As we mentioned earlier, any failure to meet timing requirements disrupts the stability of the system and can be catastrophic. For instance, consider the rover example from Sec.~\ref{sec:prob_overview}. If the controller task fails to issue navigation commands before the time limit, the rover may stall, or worse, may not be able to steer properly and even crash. Hence, delays caused by the checking mechanisms can also cause such problems. Table~\ref{tab:eval_plat_overview} presents additional examples for the possible consequences when tasks are unable to meet their timing requirements.

Existing work~\cite{mukherjee2019optimized,amacher2019performance,liu2018rt} shows that although TEEs are implemented on hardware, they can still cause significant overheads --- this is particularly acute in real-time applications.
For instance, consider the Linux-based TrustZone port, OP-TEE~\cite{optee} supported on
many embedded platforms. Our experiments show that the overhead of switching between normal to trusted mode is around $66$ ms for OP-TEE on a Raspberry Pi platform. For completeness, we also performed experiments on an ARMv8-M Cortex-M33 architecture using ARM FVP libraries~\cite{arm_fvp} where the regular applications were running on FreeRTOS~\cite{free_rtos} and trusted mode codes were executed on bare-metal. We find that the mode switching delays in this setup are $2$ ms. We note that the delays are higher in the Linux environment due to extra overheads (\ie execution of sequence of API calls~\cite{mukherjee2019optimized}) imposed by the Linux kernel and OP-TEE secure OS.
Although the overheads of secure calls (SMC) for switching between regular and trusted modes are platform-specific, it may still not be feasible to check multiple commands while meeting real-time guarantees. 
For example, if a task operates at $50$ Hz (\ie it is required to finish before $20$ ms)~\cite{cy_scheduleak_rtas19} and regular computation takes $10$ ms, the FreeRTOS-based setup allows at most $5$ checks in order to comply with timing requirements. Likewise, for the applications running on a Linux and OP-TEE-based Raspberry Pi platform\footnote{We implemented \pnametee using OP-TEE/Linux-based  Raspberry Pi platform to support multiple off-the-shelf systems.} (Sec.~\ref{sec:cs_experiments}), we can check at most $3$ commands per instance if the controller task operates at $5$ Hz. We therefore need smart techniques, say where \textit{only a subset of commands are vetted} while maintaining security guarantees, to support both security and performance for real-time applications. We now present my methods to achieve this (based on game-theoretic analysis) in the following section.


\section{Game-Theoretic Analysis for Smart Actuation Checks in \pnametee} \label{sec:overload_control}

We now propose a mechanism to deal with this issue of monitoring overheads. We consider the case when there exists a task $\tau_i$ such that 
it cannot perform all the $N_i$ checks before its deadline (denoted by $D_i$). 
One option to reduce the number of checks is to verify only a subset of commands so that the task can finish before its deadline. That is, check a subset of commands, $K_i$, ($N_i^{min} \leq K_i < N_i$) such that $R_i^{TEE} \leq D_i, \forall \tau_i \in \Gamma$ where $R_i^{TEE}$ is the response time (\ie time between task arrival to completion, used to verify that the task meets its deadline~\cite{res_time_rts}). 
The challenge is then to decide \textit{which} subset of $K_i$  (among $N_i$) actuation requests should be selected for checking in each task $\tau_i$. In addition, if we check only a fixed set of $K_i$ commands and an adversary jeopardizes some or all of the remaining  $N_i - K_i$ requests, then the attack will succeed and remain undetected. To balance the security and real-time requirements, \pnametee \textit{randomly selects different subsets of requests for checking}. In particular, during each task execution we randomly pick a set of $K_i$ commands with pre-computed probability distributions.
 While we pick a subset of commands, it should look like (to adversary) that \textit{\pnametee is checking all commands}. As a result, it will be difficult for an attacker to identify which subset of requests are selected for checking 
 since each instance of a task will select a different subset of requests for vetting.
As we shall see in Sec.~\ref{sec:game_forumlation} we formulate this problem as a two-player game~\cite{conitzer2006computing} and develop Algorithm~\ref{alg:multicore_act_verif} to determine the feasible number of $K_i$ inspection points that provide similar level of security when compared to the case that checks all $N_i$ requests. We now illustrate our ideas used in \pnametee with a simple example.

{\bf Intuition and Example.} Let us consider a ground rover performing a line-following mission. The rover controller task ($\tau_c$) generates the following actuation requests ($N_c = 3$): \ca $\mathtt{setEnc_L}(val)$ and $\mathtt{setEnc_R}(val)$ that set the speed of left and right motor encoders, respectively (denoted by $a_c^1$ and $a_c^2$); \cb $\mathtt{setNav}(cmd)$ that issues a navigation command where each $cmd$ specifies values to the peripheral registers for navigating the rover forward, backward, left or right directions (denoted by $a_c^3$). Recall from the description of the system model that there exists designer-provided weights $\omega_i^j$ for checking each of the commands $a_i^j$ (see Appendix~\ref{sec:sys_mod_formal}) and that is given by: $\Omega_c = \{ \omega_c^1, \omega_c^2, \omega_c^3\}$. As we shall see in Sec.~\ref{sec:game_forumlation}, the weights are used to determine which commands should be checked more often. For example, if $\omega_c^3=2$ and $\omega_c^1=\omega_c^2=2$, then \pnametee tends to check $\mathtt{setEnc_R}(val)$ twice as often as the other two commands. The checking for $a_c^1$ and $a_c^2$ is whether the speed value is within a given bound (\eg $val \in [v^-, v^+]$) and for $a_c^3$ the checking module verifies if the  $cmd$ value is consistent so that the rover is on the line and is correctly following the mission. 

\begin{figure}[!t]
	\centering
	\includegraphics[scale=0.63]{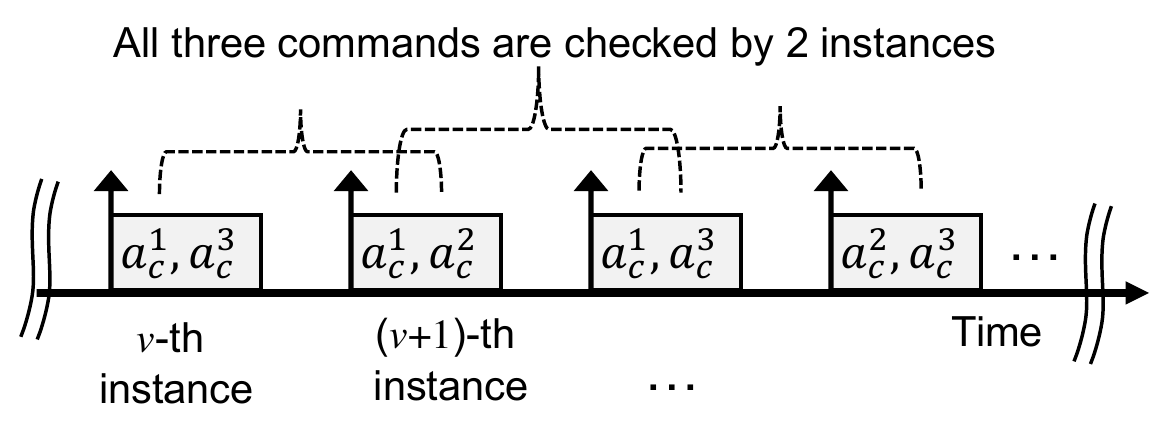}
	\caption{Non-deterministic checking: 
		\textit{all} commands are eventually checked.}
	\label{fig:check_all_timeline}
\end{figure}

We now consider the case when checking all three requests does not comply with the timing requirement of task $\tau_c$
and we can only verify at most $K_i = 2$ requests (we describe how to calculate the value of $K_i$ for each task $\tau_i$ in Sec.~\ref{sec:cal_ki}). Therefore, the possible combinations for checking are as follows: $X_c = \{ (a_c^1, a_c^2), (a_c^2, a_c^3), (a_c^1, a_c^3)  \}$. In \pnametee, for each instance of the task $\tau_c$ we randomly select any $j$-th element from the set $X_c$ with probability $x_c^j$ that provides better ``monitoring coverage''. For example, let $x_c^1=x_c^2=0.25$ and $x_c^3=0.5$. Then, for any given instance of $\tau_c$ the possibility of verifying both $a_c^1$ and $a_c^2$ is $25\%$, verifying both $a_c^2$ and $a_c^3$ is $25\%$, where the possibility of verifying $a_c^1$ and $a_c^3$ is $50\%$ (recall that by assumption we can check only $2$ commands per job). Figure~\ref{fig:check_all_timeline} presents an instance of such execution. \pnametee ensures that although we pick a subset of commands each time, eventually \textit{all} the commands are checked. For example, in Fig.~\ref{fig:check_all_timeline} \pnametee requires two instances to check all three commands. In the following section we present our ideas to compute these probabilities using game-theoretic analysis.


\subsection{Generating Randomized Schedules} \label{sec:game_forumlation}

We now present our techniques to derive the probabilities for selecting subsets of commands to be checked. The formulations this section assumes that the size of feasible subset of  $K_i$ commands, that ensures all timing requirements are met, is known for each task. In Sec.~\ref{sec:cal_ki} we present algorithms to derive $K_i$. We model the selection of a subset of commands (for checking) as a \textit{two-player normal-form game} (also called leader-follower game)~\cite{harsanyi1972generalized, conitzer2006computing, paruchuri2007efficient}.
The game formulations allow the designers to model the fact that an attacker acts with knowledge of the defender's actions and react accordingly. Since normal-form games  address  the  challenges  posed in  our  context (\ie selecting optimal actions for decision-making agents),  we use this game model in \pnametee for generating randomized schedules.

\begin{figure}[!t] 
	\hspace*{-0em}\underline{\em System Reward}:
	
	\begin{equation}\label{eq:system_reward}
		\hspace*{-7em}
		\lambda_{i}^{j,l} = \frac{\sum\limits_{w \in \tikzmarknode{psix}{\highlight{blue}{\Psi(X_i^j)} }} \tikzmarknode{wreward1} {\highlight{olive}{w}} }{\sum\limits_{w \in \tikzmarknode{cup1}{\highlight{violet}{ \Psi(X_i^j) \cup \Psi(Q_i^l) }} }  \tikzmarknode{wreward2}{ \highlight{olive}{w}   }}, 
	\end{equation}
	\vspace*{0.8\baselineskip}
	\begin{tikzpicture}[overlay,remember picture,>=stealth,nodes={align=left,inner ysep=1pt},<-]
		\path (psix.north) ++ (0.1,2.2em) node[anchor=north west,color=blue!85] (psixtext){\footnotesize weights in designer's $j$-th strategy};
		\draw [color=blue](psix.north) ++ (0.1, 0) |- ([xshift=-0.3ex,color=blue]psixtext.north east);
		\path (wreward1.north) ++ (3.2,-1.0em) node[anchor=north east,color=xkcdDarkOlive!85] (wreward1text){\footnotesize summing up weights};
		\draw [color=olive](wreward1.south) |- ([xshift=-0.3ex,color=olive]wreward1text.south east);
		\path (cup1.north) ++ (-3.7,-2.6em) node[anchor=north west,color=violet!85] (cup1text){\footnotesize normalization factor: union of designer's and adversary's strategies};
		\draw [color=violet](cup1.south) |- ([xshift=-0.3ex, yshift=2.7ex,color=violet]cup1text.south west);
		\path (wreward2.north) ++ (3.2,-1.0em) node[anchor=north east,color=xkcdDarkOlive!85] (wreward2text){\footnotesize summing up weights};
		\draw [color=olive](wreward2.south) |- ([xshift=-0.3ex,color=olive]wreward2text.south east);
	\end{tikzpicture}

	\hspace*{-00em}\underline{\em System Cost}:
	
	\begin{equation} \label{eq:system_cost}
		\hspace*{-7em}
		\zeta_{i}^{j,l} = \frac{\sum\limits_{w \in \tikzmarknode{zetax}{\highlight{blue}{\Psi(Q_i^l)} }} \tikzmarknode{wreward1} {\highlight{olive}{w}} }{\sum\limits_{w \in \tikzmarknode{cup1}{\highlight{violet}{ \Psi(X_i^j) \cup \Psi(Q_i^l) }} }  \tikzmarknode{wreward2}{ \highlight{olive}{w}   }}, 
	\end{equation}
	\vspace*{0.8\baselineskip}
	\begin{tikzpicture}[overlay,remember picture,>=stealth,nodes={align=left,inner ysep=1pt},<-]
		\path (zetax.north) ++ (0.1,2.2em) node[anchor=north west,color=blue!85] (zetaxtext){\footnotesize weights in adversary's $l$-th strategy};
		\draw [color=blue](zetax.north) ++ (0.1, 0) |- ([xshift=-0.3ex,color=blue]zetaxtext.north east);
		\path (wreward1.north) ++ (3.2,-1.0em) node[anchor=north east,color=xkcdDarkOlive!85] (wreward1text){\footnotesize summing up weights};
		\draw [color=olive](wreward1.south) |- ([xshift=-0.3ex,color=olive]wreward1text.south east);
		\path (cup1.north) ++ (-3.7,-2.6em) node[anchor=north west,color=violet!99] (cup1text){\footnotesize normalization factor: union of designer's and adversary's strategies};
		\draw [color=violet](cup1.south) |- ([xshift=-0.3ex, yshift=2.7ex,color=violet]cup1text.south west);
		\path (wreward2.north) ++ (3.2,-1.0em) node[anchor=north east,color=xkcdDarkOlive!85] (wreward2text){\footnotesize summing up weights};
		\draw [color=olive](wreward2.south) |- ([xshift=-0.3ex,color=olive]wreward2text.south east);
	\end{tikzpicture}
\caption{Reward and cost functions.}
\vspace*{1em}
\label{fig:reward_cost}
\end{figure}

%

\subsubsection{Game Setup}


In our model we consider two players: The leader (\ie system designers) and the follower (\ie adversary). Let $X_i$ denote the set of all combinations of choosing $K_i$ subset of commands from total $N_i$ number of possibilities, \ie the size of set $|X_i| = {N_i \choose K_i} = \tfrac{N_i!}{K_i!(N_i-K_i)!}$. In game-theoretic terminology $X_i$ represents the set of leader's ``strategies''. As discussed in the previous section, the $j$-th element of $X_i$ is a vector of size $K_i$ that represents which subset of commands will be picked for inspection in the secure enclave.  Let us now introduce the variable $Q_i$ that represents the attacker's set of actions (\ie follower's strategies). The set $Q_i$ represents the possible combinations of actuation requests invoked by task $\tau_i$ that can be compromised by an adversary. Recall from the rover example where the controller task $\tau_c$ invokes $N_c=3$ actuation requests, hence, the adversary can pick one of the following eight combinations: $Q_c = \{ (a_c^1), (a_c^2), (a_c^3), (a_c^1, a_c^2), (a_c^2, a_c^3), (a_c^1, a_c^3), (a_c^1, a_c^2, a_c^3), (\emptyset) \}$. For example, the first element in the set denotes the adversary chooses to compromise only invocation $a_c^1$, the fifth element implies both $a_c^2$ and  $a_c^3$ are compromised while the last element implies there is no attack during this instance of the task. Note that the size of the attacker's strategy set $Q_i$ is $2^{N_i}$. 

Recall that each actuation command $a_i^j$ is associated with a designer-given weight $\omega_i^j$ (see Sec.~\ref{sec:sys_mod} and Appendix~\ref{sec:sys_mod_formal}). A higher weight implies that designers want to check the corresponding command more often. For instance, from the rover example, designers may want to check navigation commands ($a_c^3$) more frequently than the ones that set the wheel speeds ($a_c^1$, $a_c^2$) and may set higher weight for $\omega_c^3$. Let $\Lambda(X_i^j)$ denote the set of commands used for vetting and $\Psi(X_i^j)$ is the set of corresponding weights in the $j$-th element of the strategy set $X_i$. Likewise $\Lambda(Q_i^l)$ denotes the set of commands and $\Psi(Q_i^l)$ is the corresponding set of weights compromised by the attacker in its $l$-th strategy. In the rover example, if we select $j=2$ and $l=4$ (\ie second and fourth elements of the designers and adversary's strategy set) then $\Lambda(X_c^2) = \{ a_c^2, a_c^3 \}$, $\Psi(X_c^2) = \{ \omega_c^2, \omega_c^3 \}$ and $\Lambda(Q_c^4) = \{ a_c^1, a_c^2\}$, $\Psi(Q_c^4) = \{ \omega_c^1, \omega_c^2\}$. 

We now introduce two variables, \viz \textit{system reward} ($\lambda$) and \textit{system cost} ($\zeta$). A \textit{higher system reward and lower cost} is \textit{good for the designers} and \textit{bad for the attackers}. Likewise, higher system cost and lower reward is favorable for the adversary's point of view (and bad for the designers). If a task $\tau_i$ selects the $j$-th element from the set of strategies $X_i$ and the attacker selects the $l$-th strategy from $Q_i$ for attack then the system reward is $\lambda_i^{j,l}$ and cost is $\zeta_i^{j,l}$. If the task selects a subset of commands for vetting in its $j$-th strategy and the adversary also attacks those invocations in its $l$-th strategy, \ie $\Lambda(X_i^j) = \Lambda(Q_i^l)$, it implies that the attack is detected. Hence, we set $\lambda_i^{j,l}$  to a large positive value (\ie high system reward, since the attack is detected) and $\zeta_i^{j,l}$ a large negative value (\ie no system cost). In contrast, if $\Lambda(X_i^j) \cap \Lambda(Q_i^l) = \emptyset$ for any pair $(j, l)$, \ie  $\Lambda(X_i^j)$ does not contain any commands in attackers $l$-th strategy $\Lambda(Q_i^l)$, that implies the compromised commands are not vetted (\ie the spoofed command is not checked). In this case we set  $\lambda_i^{j,l}$ a large negative value (\ie no reward) and $\zeta_i^{j,l}$ a large positive value (\ie high system cost). When the above two conditions do not hold (\ie only a subset of the compromised commands are checked) and therefore $\exists (j,l)$ such that $\Lambda(X_i^j) \cap \Lambda(Q_i^l) \neq \emptyset$, we then obtain the system reward/cost by \textit{normalizing the weights} of both adversary and designer's strategies. For this, we define the reward and cost functions in Eq.~(\ref{eq:system_reward}) and Eq.~(\ref{eq:system_cost}), respectively (see Fig.~\ref{fig:reward_cost}). Let us revisit the rover example with $j=2$ and $l=4$. In this case $\lambda_{c}^{2,4} = \tfrac{\omega_c^2 + \omega_c^3}{\omega_c^1 + \omega_c^2 + \omega_c^3}$ and $\zeta_{c}^{2,4} = \tfrac{\omega_c^1 + \omega_c^2}{\omega_c^1 + \omega_c^2 + \omega_c^3}$.


This reward and cost functions give us one way to measure the security of the system in terms of how many significant actuation commands we can monitor given an attacker's strategy. A higher system reward (and lower cost) implies that \pnametee performs more checking with respect to a given adversarial action.

\subsubsection{Formulation as an Optimization Problem}

We now develop models to determine the optimal strategy for each of the tasks. Let us now denote $x_i^j$ as the probability of selecting the $j$-th element from $X_i$ (represents  the  proportion  of  times in  which a strategy $j$ is used by the task $\tau_i$ in the game). The output of the game will provide the probability distribution of (randomly) selecting subset of $K_i$ commands from the set possible choices (\ie $X_i$) for the different execution instances of a given task $\tau_i$. For a given adversarial strategy $l$, 
summing over all the strategy sets $X_i$ (\ie $\sum\nolimits_{j=1}^{|X_i|} x_i^j \lambda_i^{j,l}$ and $\sum\nolimits_{j=1}^{|X_i|} x_i^j \zeta_i^{j,l}$) gives us the total system reward and cost, respectively. 

We can obtain probability distributions of selecting elements from $X_i$ for a given attacker strategy $l$ (that maximizes the system reward) by forming a linear optimization program. In particular, 
for each of the attacker's $l$-th strategy (where $1 \leq l \leq |Q_i|$), we compute a
strategy for the $\tau_i$ such that \ci playing $l$-th strategy is a best response from the adversary's point of view (\ie more system cost) and \cii under this constraint, the strategy maximizes the reward for $\tau_i$ (\ie checks critical commands more often). Appendix~\ref{sec:game_linear_formulation} presents the details of our linear programming formulation.

\subsection{Calculating the Size of the Feasible Command Set} \label{sec:cal_ki}

Our focus here is to \textit{examine as many actuation commands as possible} while meeting real-time guarantees. The game formulation from the previous sections assumes that we know the size of the set $K_i$ and calculate the probabilities accordingly. However, in a system with multiple real-time tasks, finding the size of feasible set $K_i$ for each task $\tau_i \in \Gamma$ while also meeting the real-time requirements (deadlines) is a non-trivial problem. We therefore develop an iterative solution for finding the size of this set.

\renewcommand{\algorithmicforall}{\textbf{for each}}
\renewcommand\algorithmiccomment[1]{%
 {\it /* {#1} */} %
}
\renewcommand{\algorithmicrequire}{\textbf{Input:}}
\renewcommand{\algorithmicensure}{\textbf{Output:}}
    
\begin{algorithm}[t]
    \begin{algorithmic}[1]
    	\begin{footnotesize}
        	\REQUIRE Input taskset parameters $\Gamma$
        
            \ENSURE For each task $\tau_i$, the size of the feasible set $K_i^* \geq N_i^{min}$ and selection probability $x_i^j, j=1, \cdots, |X_i^*|$ for each of the combinations in the strategy set $X_i^*$; $\mathsf{Infeasible}$   otherwise.
            
            \STATE \COMMENT{Check minimum feasibility requirements}
            \FORALL{ $\tau_i \in \Gamma$}
                \STATE Set $K_i = N_i^{min}$ and calculate response time $R_i^{TEE}$ using Eq.~(\ref{eq:tz_rta})
            \ENDFOR
        	
            \STATE \COMMENT{Unable to integrate {\normalfont \pnametee} with minimum QoS requirements}
            \IF{$\exists \tau_i$ such that $R_i^{TEE} > D_i$}
                \RETURN$\mathsf{Infeasible}$  
            \ENDIF
            
                \FORALL{task $\tau_i$ (from higher to lower priority order)}
                
                    \STATE Find maximum $K_i^* \in [N_i^{min}, N_i]$ such that all low-priority tasks $\tau_l$ meet their timing requirements (\ie $R_l^{TEE} \leq D_l$) 
                    
                    \STATE \COMMENT{not all the commands can be examined -- obtain parameters for non-deterministic checking}
                    \IF{$K_i^* < N_i$}
                        \STATE Determine the strategy set $X_i^*$ for $K_i^*$ where $|X_i^*| = {N_i \choose K_i^*}$ and obtain probabilities $x_i^j$ by solving the 
                        game formulation
                    \ENDIF
                    
                    \STATE Update response time $R_l^{TEE}$ for each $\tau_l$ that executes with a priority lower than $\tau_i$  with the updated size $K_i^*$ 
                
                \ENDFOR
                
            
            \STATE \COMMENT{return the solution}
            \RETURN the size of the feasible set $K_i^*$ and probability $x_i^j$ of selecting $j$-th strategy ($j=1,2,\cdots, |X_i^*|$) from $X_i^*$ for each task $\tau_i \in \Gamma$

            \end{footnotesize}
	\end{algorithmic}
    
\caption{\pnametee: Parameter Selection}
\label{alg:multicore_act_verif}
\end{algorithm}

Our proposed solution works as follows  (refer to Algorithm~\ref{alg:multicore_act_verif} for a formal description).  In Lines 1--4, we first assign $K_i = N_i^{min}, \forall \tau_i$ and check whether all tasks meet their timing requirements (\ie finish before their deadlines). If there exists a task that fails to meet its timing requirements, we report that it is ``\textit{infeasible}'' to integrate \pnametee in the target system while still satisfying designer specified QoS requirements (Line 7). This infeasibility result provides hints to the designers to either update or modify system parameters (\eg number of commands, QoS requirements) to enable the ability to check actuation commands in the system. Otherwise, 
we optimize the number of commands a task can verify in an iterative manner (Lines 9--16). To be specific, for a given  task $\tau_s$ 
we perform a logarithmic search 
(see Algorithm~\ref{alg:act_verif_log_search} in Appendix~\ref{sec:alg_bin_search_appsec} for the pseudo-code) and find the maximum number of commands $K_i^*$ that can be verified within the range $[N_i^{min}, N_i]$ such that all low-priority tasks $\tau_l$ meets their timing requirements with actuation checks enabled (Line 10).  
If the selected parameter $K_i^*$ is less than the total commands $N_i$, we then use game theoretical-analysis from Sec.~\ref{sec:game_forumlation} and obtain probabilities of randomly selecting  $K_i^*$ commands (in each task instance) from a total of $N_i$ commands (Line 13). The above process is repeated for all the tasks.

\section{Evaluation} \label{sec:eval}

In this section we first present our implementation details (Sec.~\ref{sec:implementation}). We then show the viability of \pnametee using \ci~four realistic cyber-physical case-studies (Sec.~\ref{sec:cs_experiments}) and \cii simulated workloads for a broader design-space exploration (Sec.~\ref{sec:syn_experiments}). 

\begin{table}
	\caption{Summary of Our Implementation Platform}
	\label{tab:imp_plat}
	\centering
	\begin{tabular}{P{2.2cm}||P{5.30cm}}
		\hline 
		\bfseries Artifact & \bfseries Configuration\\
		\hline\hline
		Platform         & Broadcom BCM2837 (Raspberry Pi 3)  \\
		CPU         & 1.2 GHz 64-bit ARM Cortex-A53  \\
		Memory & 1 Gigabyte  \\ 
		Operating System & Linux (NW), OP-TEE (SW)							\\ 
		Kernel version   & Linux kernel 4.16.56,
		OP-TEE core 3.4  \\ 
		Interface & I2C \\
		Boot parameters & \texttt{dtparam=i2c\_arm=on},     \texttt{dtparam=spi=on},
		\texttt{force\_turbo=1}, \texttt{arm\_freq=1200},
		\texttt{arm\_freq\_min=1200}, \texttt{arm\_freq\_max=1200} \\
		\hline
	\end{tabular}
\end{table}

\subsection{Implementation} \label{sec:implementation}

We implemented 
\pnametee on Raspberry Pi 3 Model B~\cite{rpi3} (equipped with 1.2 GHz 64-bit ARMv8 CPU and 1 GB RAM). We selected Raspberry Pi as our implementation platform since \ca it is a COTS system that supports a commodity TEE (ARM
TrustZone), \cb existing literature has shown the feasibility of deploying
cyber-physical applications on Raspberry Pi~\cite{securecore_syscal, cheng2017orpheus, virtsense_liu2018, protc_liu2017, liu2018alidrone,mhasan_iotsnp19} and \cc it provides a robust development environment that allows us to analyze the viability of our approach for multiple realistic off-the-shelf cyber-physical systems on a common platform. In our experiments we considered both, motors (DC as well as stepper) and servos as actuators. We used the Adafruit motor shield~\cite{adafruit_motor_hat} (an I/O extension daughter-board for Raspberry Pi) that allowed us to control multiple actuators using the I2C interface. We used an open-source motor driver~\cite{adafruit_motor_hat_code} and servo controllers~\cite{pca9685}. We implemented the trusted execution modes using the OP-TEE~\cite{optee} software stack that uses GlobalPlatform TEE APIs~\cite{globalplatform_tee_api}. 
OP-TEE provides a minimal secure kernel (called OP-TEE core) that can be run in parallel with a rich OS (\eg Linux). We used an Ubuntu 18.04 filesystem with a  64-bit Linux kernel (version 4.16.56) as the rich OS and executed our \checkact functions in the OP-TEE secure kernel (version 3.4). Our controller and checker codes are written in C for compatibility with the OP-TEE APIs. For accuracy of our measurements we disabled all the frequency scaling features in the kernel and executed RP3 at a constant frequency (\ie 1.2 GHz, the maximum supported clock speed). This was to ensure that values observed in different trials and case-studies were consistent.

We solved the linear programs using the Python-MIP library~\cite{python_mip} with CBC solver~\cite{cbc_solver}. From the probabilities obtained by the game model (Sec.~\ref{sec:game_forumlation}), at runtime (\ie for each instance of a task) we used the roulette-wheel selection technique~\cite{lipowski2012roulette} and a standard C random number generator for selecting a random subset of commands. However, this does not preclude the use of other hardware-supported generators such as Z1FFER~\cite{Z1FFER} and OneRNG~\cite{OneRNG} to ensure tamper-proof true random number generation that will further improve the security of \pnametee. For each of the selected commands, control was transferred to the secure enclave (\ie OP-TEE core) for the checking. For each of our case-studies, we implemented the \checkact functions as OP-TEE trusted applications. 
The implementation and details of \checkact for the each of platforms are presented in Sec.~\ref{sec:cs_experiments}. We note that our implementation using Raspberry Pi, Linux and OP-TEE serves as a good proof-of-concept and can be extended with other OS, hardware platforms and TEE architectures without loss of generality. Our implementation code is available in a public repository~\cite{rt_act_sec_game_repo}. Table \ref{tab:imp_plat} summarizes the system configurations and implementation details.


\begin{table*}[!t]
\caption{Real-Time Cyber-Physical Platforms used in Our Experiments}
\label{tab:eval_plat_overview}
\hspace*{-0.5em}
\centering
\scriptsize
\begin{tabular}{P{1.9cm}||P{3.0cm}|P{2.5cm}|P{2.5cm}|P{2.5cm}|P{3.0cm}}
\hline 
\bfseries Platform & \bfseries Application & \bfseries Real-Time/Safety Requirements & \bfseries  Actuation Commands & \bfseries Attack Demonstration & \bfseries Checks inside Enclave \\
\hline\hline
Ground Rover
\begin{minipage}{.3\textwidth}
\includegraphics[scale=0.08]{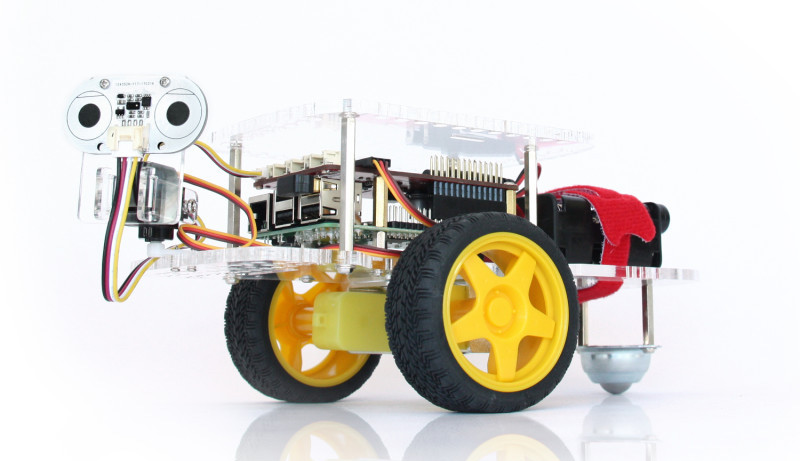}
\end{minipage}
& The rover performs a line following mission. The controller task sets the speed of the rover and steers the wheels (based on its position on the line) by executing a PID control loop  & Set the speed and direction of the motors for the  wheel movements within sampling interval (\ie control loop frequency, set at $5$ Hz) & \textbullet~Set the speed of the wheels \par \textbullet~Set wheel directions (left, right, forward and backward)  & DoS attack~\cite{mhasan_iotsnp19}: arbitrarily sets high speed for one of the wheel motors  & The speed of motors can only be within predefined limit  \\
\hline
Flight Controller
\begin{minipage}{.3\textwidth}
\includegraphics[scale=0.05]{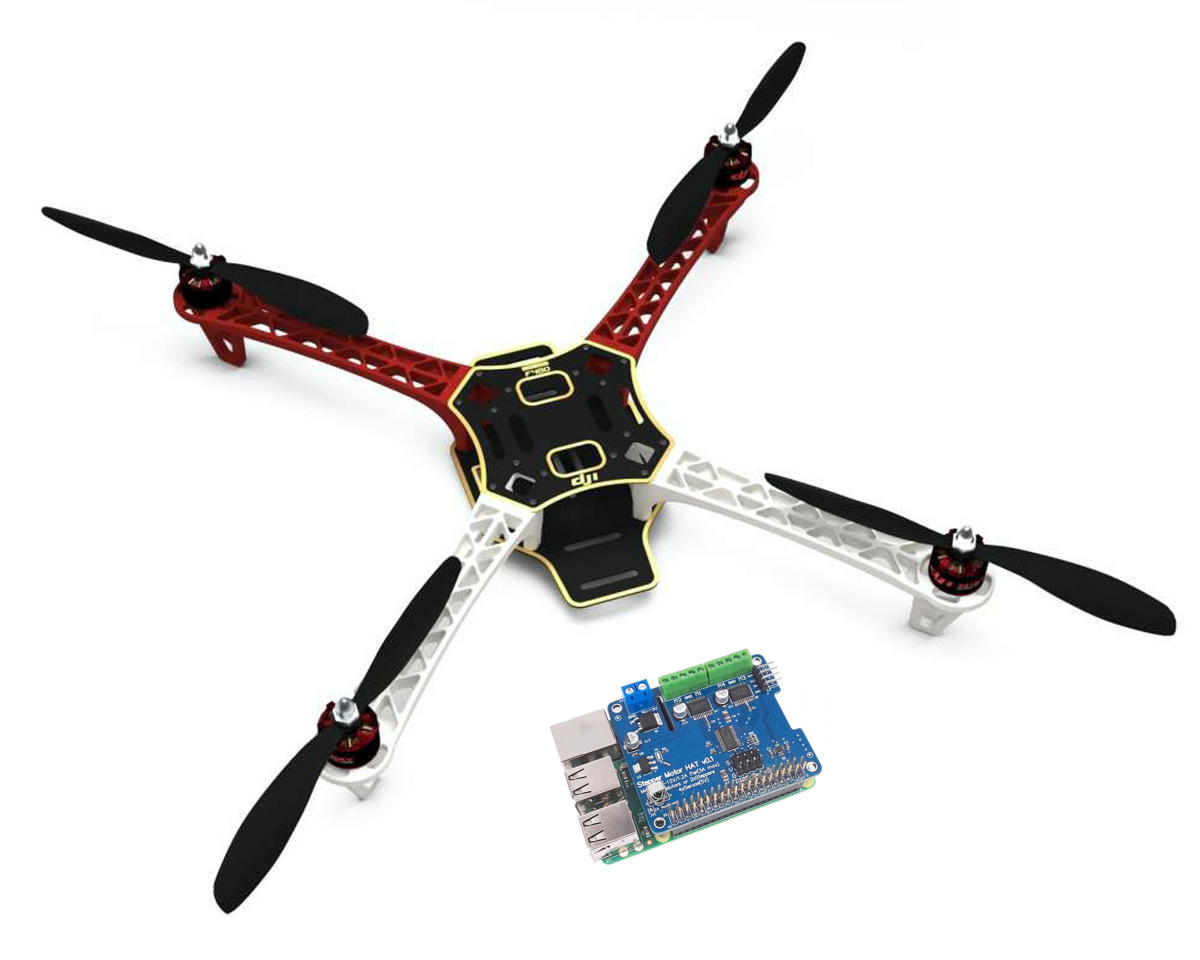}
\end{minipage}
 & Executes a PID control loop and issues PWM signals to four motors connected to the four propellers of a quad-copter  & Issue the PWM signal within sampling frequency interval ($5$ Hz in our setup) to ensure the quad-copter is stable & \textbullet~Set PWM frequency \par \textbullet~Set PWM pulse duration (four, one for each of the propellers)  & Parameter corruption attack~\cite{choi2018detecting}: modify PID control coefficients and send incorrect PWM pulse to the front right motor & Check PID control coefficients (\ie pulse duration values) before issuing PWM signals to the motos  \\
 \hline
 Robotic Arm
\begin{minipage}{.3\textwidth}
\includegraphics[scale=0.1]{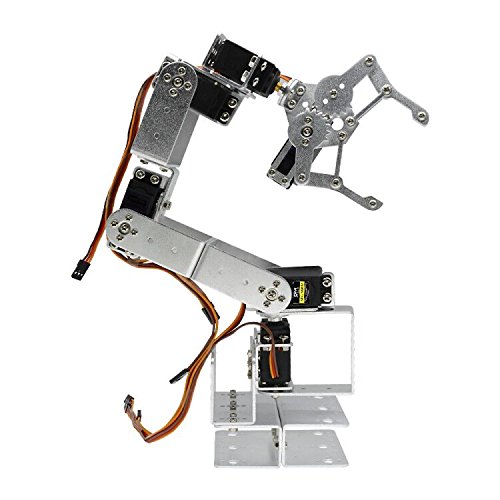}
\end{minipage}
 & The robot arm performs the following operations in a sequence: pick an object (close it claws), move the arm to destination position, drop the object (open claws) and reset the arm back to initial position & Complete movement of the object before arrival of the next object; inter-arrival duration of the objects was set at $250$~ms & Set rotation angle for each of the four servos & Synchronization attack~\cite{artinali}: sends incorrect \texttt{angle} value to the servo channel and prevents the arm from resetting back to its initial position & Check the consistency of each \texttt{(channel, angle)} pair (\ie the \texttt{angel} value for a given servo \texttt{channel} can not be more than the designer provided bounds) before issuing pulses to the servo motors \\
  \hline
 Syringe Pump
\begin{minipage}{.3\textwidth}
\includegraphics[scale=0.1]{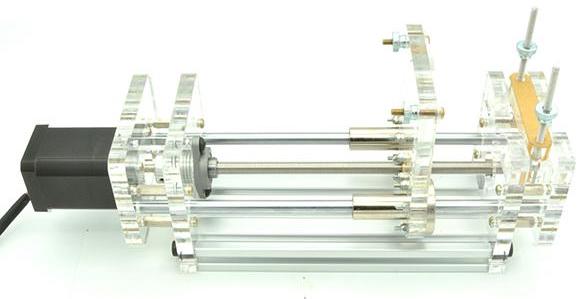}
\end{minipage}
 & The pump pushes certain amount of fluid and then pulls the trigger to reset the syringe to its initial position & Perform the push/pull operations within designer specified time limit (set at $300$ ms) & \par \textbullet~ Set motor rotation frequency \par \textbullet~Drive the motor forward/backward to push the fluids and reset the motor position  & Bolus tampering attack~\cite{cheng2017orpheus, artinali}: the attacker injects more fluid than the permitted volume & Checks the amount of fluid the motors can pump (\ie monitor the number of \texttt{PUSH}/\texttt{PULL} events the controller task invokes) \\
\hline
\end{tabular}
\end{table*}

\subsection{Experiments with Realistic Cyber-Physical Platforms} 
\label{sec:cs_experiments}

We chose four realistic real-time cyber-physical platforms as case-studies to evaluate the efficacy of \pnametee: \ca ground rover, \cb UAV flight controller, \cc robotic arm and \cd syringe/infusion pump that are used in many cyber-physical applications. 
These are \textit{off-the-shelf platforms and we did not modify them.} Table~\ref{tab:eval_plat_overview} summarizes the properties of each of these systems and attack/detection techniques used in our experiments. 
We note that unlike generic applications, there are few publicly available open-source real-time platforms due to their proprietary nature (see more in Sec.~\ref{sec:discussion}). 
In addition, there is a significant amount of effort involved in setting up a TEE-supported real-time cyber-physical platform and generating evaluation traces from it. 
We therefore limit ourselves to four real-time platforms in this paper --- albeit they cover a wide range of application domains (see Table~\ref{tab:rt_act_exp}) and should suffice to demonstrate the feasibility of our approach.

Since we focus on scheduling independent actuation checking events, the type of attacks and checking techniques are orthogonal to our model. 
For demonstration purposes, we use fault injection~\cite{tabrizi2015flexible,artinali} to mimic malicious behavior and trigger attacks that are known to the checking module (\ie \checkact  function). 
Note that this is a standard technique used by the researchers to evaluate security solutions in cyber-physical applications~\cite{choi2018detecting, securecore, guo2018roboads, mhasan_resecure_iccps,mhasan_iotsnp19,artinali}.
We now present the evaluation platforms used in our experiments. 
%



\textbullet~~\textbf{\textit{Case-Study \#1 (Ground Rover):}} 
Our first case-study platform is the ground rover introduced in Sec.~\ref{sec:prob_overview}. 
There are two motors attached to the rover wheels (the actuators in our context). 
We used an open-source implementation of the rover controller (written in Python)~\cite{rpi_rover_using_motor_hat} and ported it to C for compatibility with OP-TEE APIs. 
The rover performed a line-following mission where it moved from a source to a target way-point by following a line. 
Each instance of the controller task first set the speed of the motor for the wheel movements and then steered (\eg forward, left or right) based on its position on the line. 

\noindent \textit{\underline{Actuation:}} 
The rover has four actuation commands: two for setting the speed of both of the wheels and two for issuing navigation commands to the two motors attached to the wheel. 
Table~\ref{tab:eval_plat_overview} lists these actuation commands. 

\noindent \textit{\underline{Attack and Consequences:}} 
We injected a DoS attack~\cite{mhasan_iotsnp19} that arbitrarily sets a high speed for one of the motors (to destabilize the rover and move it away from the line). 
This attack can deviate the rover form its way-points (or even crash it) due to the imbalance in the wheel speeds. 
 
\noindent \textit{\underline{Detection:}} 
In our setup the \checkact functions validate whether the rover speed is within designer-given predefined thresholds~\cite{gpg2} and also verifies whether the navigation commands were consistent with the rover position. 
We detected the DoS attack by checking the bounds on the speed (\ie $70$--$100$ decimal values~\cite{gpg2}) issued by the controller task.

\textbullet~~\textbf{\textit{Case-Study \#2 (Flight Controller):}} 
Our second case-study is a flight-controller for a quad-copter~\cite{drone_flight_controller}. 
The original controller code is developed for Arduino platforms. 
Since Arduino boards do not support TrustZone, we ported it to the Raspberry Pi and OP-TEE enabled environments. 
In this setup the controller executes a PID control loop using the Ziegler–Nichols method~\cite{aastrom2004revisiting} and sends pulse width modulation (PWM) signals to spin each of four motors (\ie actuators) connected to the propellers. 

\noindent \textit{\underline{Actuation:}} 
There are five actuation commands: one for setting the PWM frequency and other four are for sending PWM pulse durations for each of the motors to rotate the copter propellers. 
The \checkact functions validate whether the PWM frequency and pulse durations sent to each of the motors were within a certain range (obtained from PID control logic).

\noindent \textit{\underline{Attack and Consequences:}} 
For this case study we considered a parameter corruption attack~\cite{choi2018detecting} that modifies the control parameters (\eg the PID control coefficients) at runtime and sends incorrect pulse values to the front-right motors. 
This attack can suddenly turn off/freeze the propellers. 
As a result, the copter will instantly fall/crash. 

\noindent \textit{\underline{Detection:}} 
This attack is detected since we verify the PID parameters and corresponding PWM pulse durations. 
    
\textbullet~~\textbf{\textit{Case-Study \#3 (Robotic Arm):}} 
Our next case-study platform is a robotic arm used in manufacturing systems. 
The movement of the robotic arm is controlled by four servos (actuators in our context). 
Each servo is connected to a specific ``channel'' (I/O port) in the Adafruit motor shield. 
We use an open-source Python-based robot controller~\cite{robot_arm_src} and adapted the implementation for our C-based setup. 

\noindent \textit{\underline{Actuation:}} 
Our robot performed an assembly line sequence with the following four actuation operations: \texttt{PICK()}, \texttt{MOVE()}, \texttt{DROP()} and \texttt{RESET()} that \ci picks an object from first position, \cii moves the arm to a final position, \ciii drops the object and, finally, \civ  resets the arm to initial position (to pick up another object). 
Each operation takes a \texttt{(channel, angle)} pair that controls the rotation of the corresponding servo to the desired angle ($45^{\circ}$ in our setup~\cite{robot_arm_src}). 

\noindent \textit{\underline{Attack and Consequences:}} We used a synchronization attack~\cite{artinali} that destabilizes the assembly line by preventing the robot from resetting its arm back to the initial position. 
To demonstrate this, we injected a logic bomb that sets an incorrect \texttt{angle} value in the \texttt{RESET()} operation (\eg servo channel 3). 
This attack can collapse the whole assembly line since the arm is not returned to the initial position and hence is unable to pick up objects queued in the line.  

\noindent \textit{\underline{Detection:}} \checkact detects this attack since it asserts that each servo can only move up to a certain designer provided angle ($45^{\circ}$) for each of the operations.

\textbullet~~\textbf{\textit{Case-Study \#4 (Syringe Pump):}} Our final case-study platform is a syringe/infusion pump~\cite{c-flat_impl}. 
In our experiments we considered a bolus delivery use-case~\cite{cheng2017orpheus,abera2016c} where the syringe pump first pushes a certain amount of fluid (\texttt{PUSH} event) and then pulls the trigger back (\texttt{PULL} event). The syringe movement is controlled by a stepper motor. Since the original implementation is for Ardiuno platforms (and does not support TrustZone), we modified the codes to make it compatible with Raspberry Pi and its motor driver library. 

\noindent \textit{\underline{Actuation:}} For a given fluid amount, the syringe pump implementation  selects the number of steps where the \texttt{PUSH} and \texttt{PULL} events should be called. We considered each of the \texttt{PUSH}/\texttt{PULL} events as actuation requests since they set the direction of motor rotation.  In our setup there were seven actuation requests: one for setting the motor rotation frequency and six for \texttt{PUSH} and \texttt{PULL} events (three each).  \texttt{PULL} events were called after all three \texttt{PUSH} operations were completed.

\noindent \textit{\underline{Attack and Consequences:}} 
We implemented a bolus tampering attack~\cite{cheng2017orpheus, artinali} where the adversary injects more fluid than is required (\ie more that three \texttt{PUSH} events). 
The attack has serious safety consequences and is a health hazard since it can inject more fluids/medications to the patient body than the permitted amount.

\noindent \textit{\underline{Detection:}}
This attack is detected by \checkact since it verifies the motor frequency and how many times each of the \texttt{PUSH}/\texttt{PULL} events are called.

{\bf Experience and Findings.} In our study we compare \pnametee 
against a naive scheme that checks \textit{all} the actuation commands. We refer to this latter technique as the ``\textit{fine-grain}'' checking scheme. Note that in fine-grain checking, there are more context switches between the normal and secure execution modes since \textit{all} the commands are checked.
The goal of our evaluation was to \textit{study the trade-offs between security and real-time requirements}. We therefore considered the subset of commands selected for checking were no more than $50\%$ of the total number of commands so that tasks can finish before their timing requirements (see Table~\ref{tab:eval_plat_overview})\footnote{We also carried out additional experiments to show the effect of varying this parameter (see Sec.~\ref{sec:syn_experiments}).} (\ie $K = \lfloor 0.5 N \rfloor$) and assumed equal weights for all commands. We note that our implementation is modular and can be easily adjusted with different weight values. Table~\ref{tab:cs_n_act_com} lists the total number of commands ($N$) and subset of commands $(K)$ for each of evaluation platforms. We now discuss the results of applying these schemes on the four case-studies and address the following research questions (RQs):

\textbullet~\textbf{RQ1.} \textit{How quickly an intrusion can be detected by} \pnametee \textit{when compared to the fine-grain scheme?}

\textbullet~\textbf{RQ2.} \textit{What are the performance impacts and runtime overheads  of these schemes?}

\begin{table}[!t]
\caption{Number of Actuation Commands}
\label{tab:cs_n_act_com}
\centering
\begin{tabular}{P{2.00cm} || c | c}
\hline
\bfseries Platform & \bfseries Total Commands & \bfseries Selected for Vetting\\
 & \bfseries ($\boldsymbol N$) & \bfseries  ($ \boldsymbol{K = \lfloor 0.5 N \rfloor}$)\\
\hline
Ground Rover & $4$ & $2$ \\
Flight Controller & $5$ & $2$ \\
Robotic Arm & $4$ & $2$ \\
Syringe Pump & $7$ & $3$ \\
\hline
\end{tabular}
\end{table}

\begin{figure}[!t]
\centering
\includegraphics[scale=0.50]{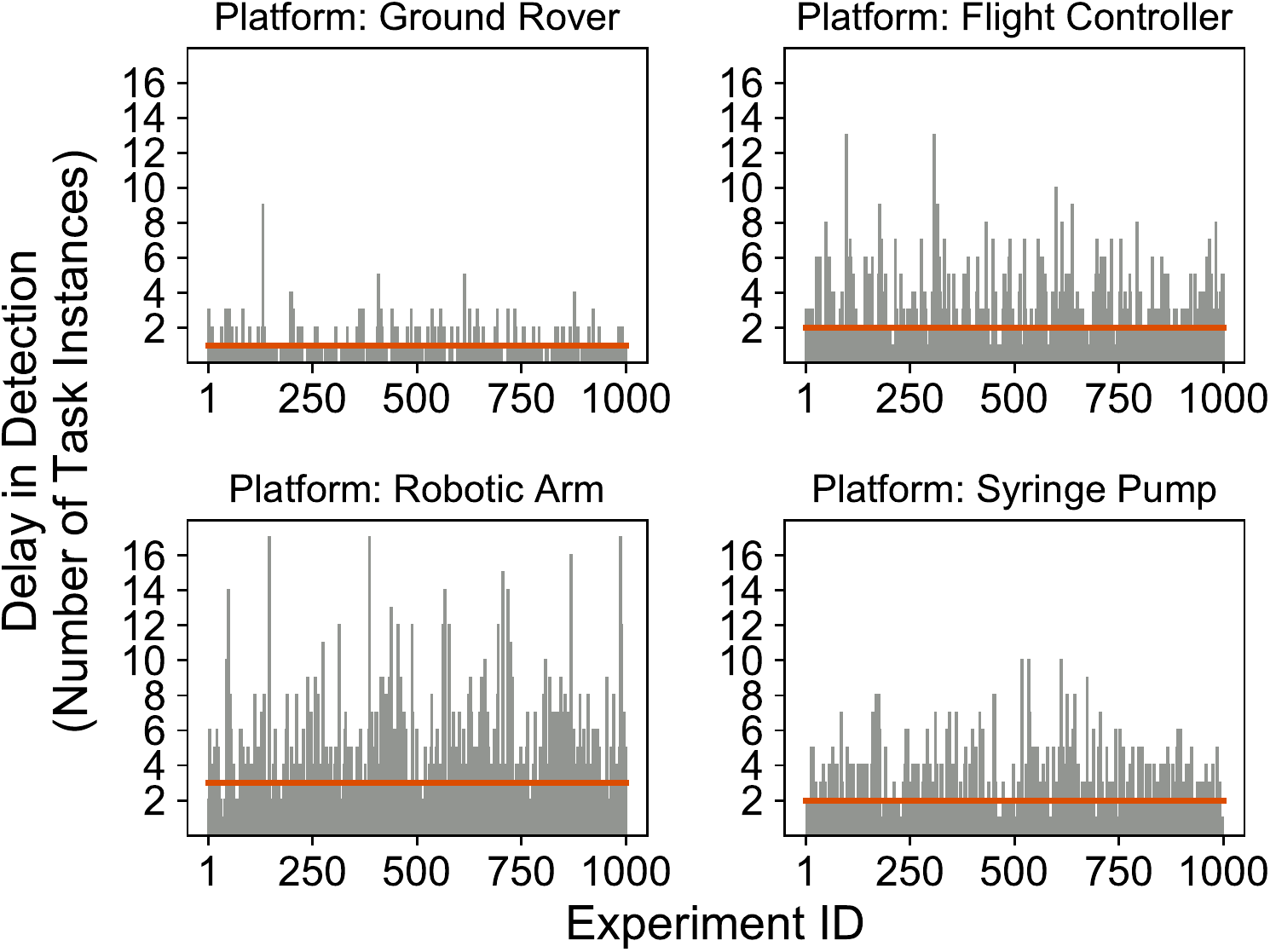}
\caption{Delay in detecting an intrusion (in terms of number of jobs) for \pnametee in comparison with fine-grain scheme. On average (orange horizontal line), the detection delay is no more than $3$ task instances for our case-study platforms.} 
\label{fig:cs_d_time}
\end{figure}

\subsubsection*{\underline{Security Analysis}} In the first set of experiments (Fig.~\ref{fig:cs_d_time}) we analyze the delay in detecting the attacks: a comparison between \pnametee and fine-grain scheme. The workflow of our experiments for each of the case-studies was as follows: for a given platform and for each of our experiments (x-axis of Fig.~\ref{fig:cs_d_time}) we triggered the attack at random points in time (\ie during the execution of the victim task) and measured the time delay (in terms of number of task instances\footnote{If the detection delay is $\hat{y}$ task instances, it implies that \pnametee requires $\hat{y}T_c$ additional time units to detect the attack when compared to the fine-grained checking where $T_c$ is the period of the corresponding controller task (see Table~\ref{tab:eval_plat_overview}).}, y-axis in Fig.~\ref{fig:cs_d_time}) in the detection of the commands by \checkact function. In the fine-grain scheme, 
the time to detect an attack is upper bounded by the sampling interval (period), $T_i$ (\ie requires at most one task instance). 
From our experiments, with $1000$ individual trials for each of the four platforms, we found that the mean and 99\textsuperscript{th}-percentile detection delay were $1$--$3$ and $3$--$12$ sampling intervals, respectively (refer to Table~\ref{tab:cs_res_summary} for exact values). We note that this delay in detection results in improve response time and reduced resource usage (see more in the following experiments) and that could be acceptable for many real-time applications\footnote{We discuss this topic further in Sec.~\ref{sec:discussion} and Appendix~\ref{sec:phys_inertia}.}.


\begin{center}
	\noindent\fbox{%
		\parbox{0.97\columnwidth}{%
			\em	\pnametee can provide similar-levels of security when compared to the fine-grained checking since, on-average, it requires only $1$--$3$ additional task instances to detect the attacks.
		}%
	}
\end{center}

\begin{figure}[!t]
\centering
\includegraphics[scale=0.50]{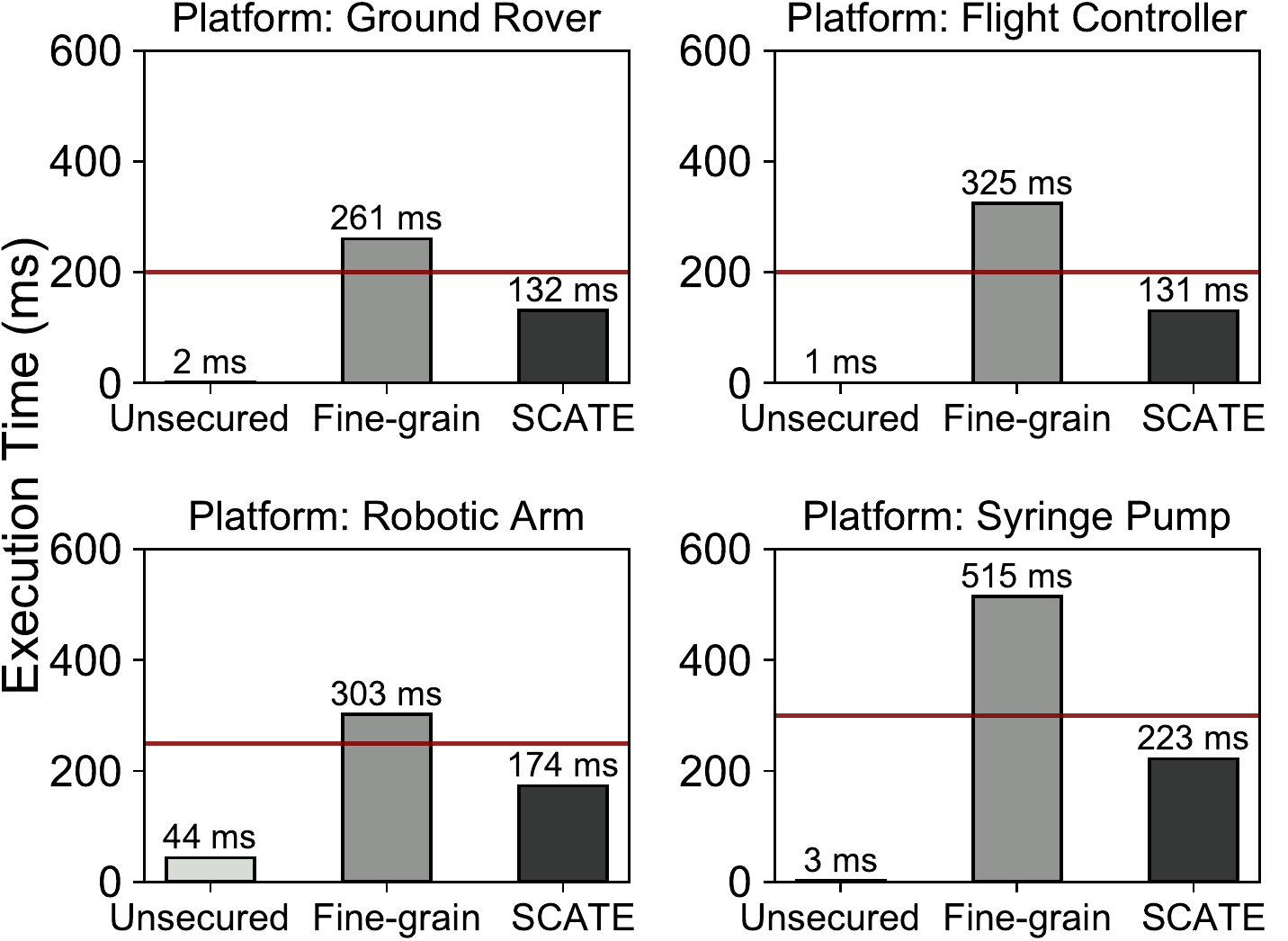}
\caption{Execution time of the main controller task for different schemes. 
The plots show 99\textsuperscript{th}-percentile values observed from $1000$ individual trials. Red horizontal lines represent task deadlines. Fine-grained checking requires more time to compute (due to additional checks and context switching overheads) and often derives the system in unsafe states.} 
\label{fig:cs_exec_time}
\end{figure}

\begin{figure}[!t]
\centering
\includegraphics[scale=0.50]{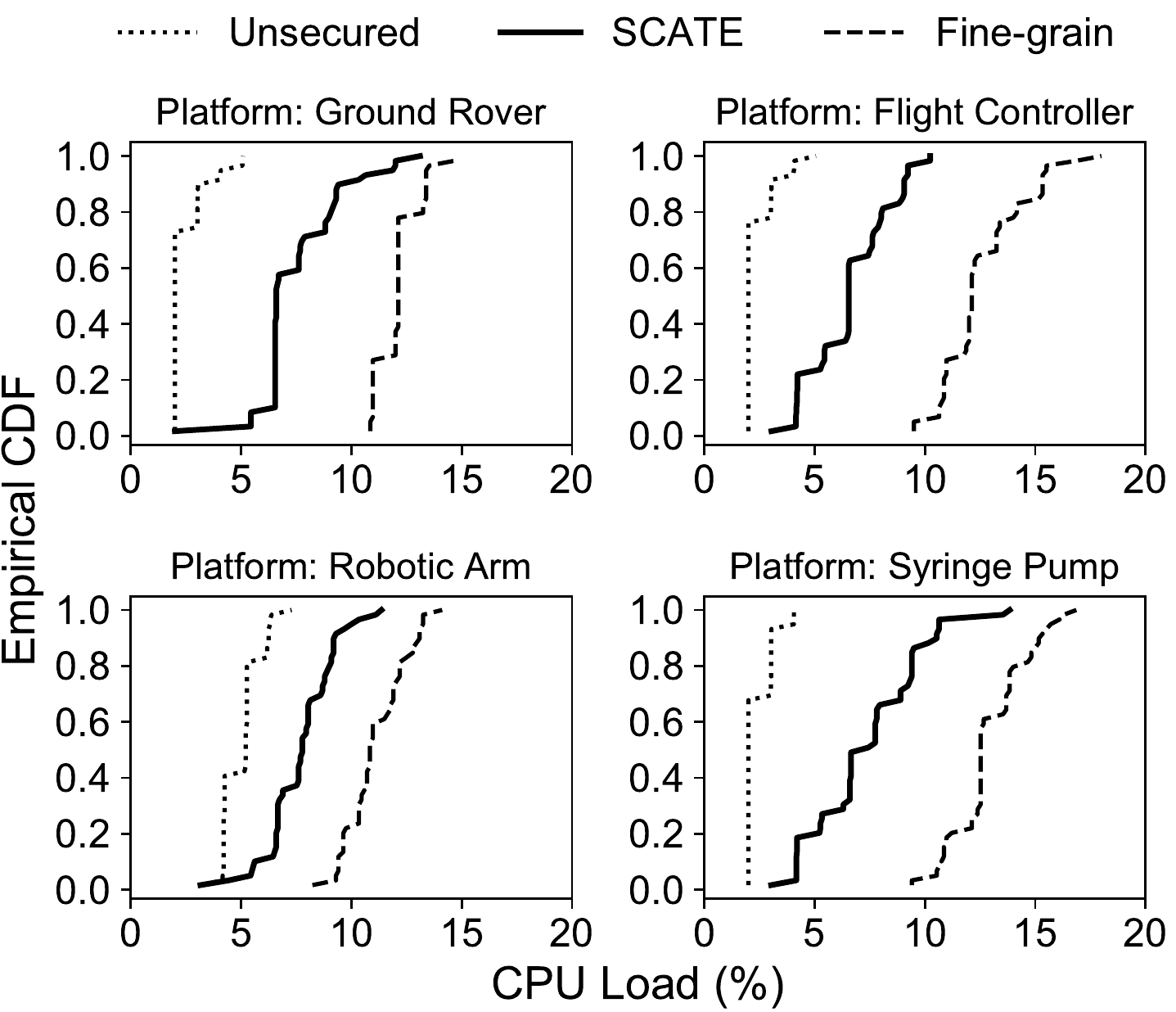}
\caption{Empirical CDF of the CPU load. 
	On average, \pnametee uses $30.48\%$--$47.32\%$ less CPU.} 
\label{fig:cs_cpu_usage}
\end{figure}

\subsubsection*{\underline{Timeliness and Overhead Analysis}}



The \textit{physical system will remain stable if the controller task can finish execution before its next (periodic) invocation}. As a reference, we also compare with traces from \textit{vanilla} execution scenarios when there is no verification of actuation commands (\ie tasks are always running in the rich OS). We refer to this vanilla execution as ``\textit{unsecured}'' since it does not protect the system from any adversarial actions.  Figure~\ref{fig:cs_exec_time} shows the execution time (y-axis) for all three schemes (captured using the Linux \texttt{clock\_gettime()} function and the \texttt{CLOCK\_PROCESS\_CPUTIME\_ID} clock). The horizontal line represents the deadlines, \ie if the response time of the task is above the margin, the task misses its deadline and the physical system will became unstable (and hence unsafe).
Both the fine-grained checking and \pnametee increase response times when compared to the unsecured scheme since there is no context switch between Linux and OP-TEE in the latter. We note that the increase in computing resources due to integration of additional security checking/protection techniques (\eg cryptographic operations, memory isolation, intrusion detection, control flow integrity checks) is an expected side-effect to improve security as observed in prior work~\cite{kim2018securing, xie2007improving, choi2018detecting, sg1, sg2, mukherjee2019optimized}.

The fine-grained scheme expends more time (when compared to \pnametee) since it verifies all $N$ actuation commands, \ie there are more context switches (from rich OS to secure enclave) and runtime checking overheads. As a result, the controller task easily misses its deadline and drive the system into an unsafe state for all of our case-studies. In contrast, intermittent checking (\pnametee) allows the tasks to finish within deadlines. We note that since, by definition, task response times must be less than their periods for stability requirements (Appendix~\ref{sec:sys_mod_formal}), the controller tasks must be required to execute with a slower frequency (\ie longer period) if we want to enforce fine-grain checking. For many control systems sampling rates affect the control performance~\cite{delay_period}; therefore by selectively checking only a subset of commands, designers can improve task response times and control frequencies while ensuring stability of the physical system. Recall from earlier discussion (Sec.~\ref{sec:intro}--\ref{sec:mot_ov_bg}) that the usefulness of real-time tasks depends on whether it finished before their deadlines (\ie earlier completion does not provide added benefits). Hence the delays due to \pnametee are acceptable since deadlines are met.


\begin{center}
	\noindent\fbox{%
		\parbox{0.97\columnwidth}{%
			\em	 Fine-grained checking increases execution times and controller tasks fail to comply with their timing requirements. \pnametee manages to complete execution before the deadline of the tasks and this is by design.
		}%
	}
\end{center}

 
In the final set of experiments (Fig.~\ref{fig:cs_cpu_usage}) we show the resource usage (\ie CPU load) for all three cases (unsecured, fine-grained and \pnametee) for each platform in our evaluation. For this, we executed the controller tasks independently for $60$ seconds and observed the CPU load using the \texttt{/proc/stat} interface. We report the results from $1000$ individual trials. 
The x-axes of Fig.~\ref{fig:cs_cpu_usage} show the CPU load and y-axes show the corresponding cumulative distribution function (CDF). 
From our experiments we found that \pnametee increases CPU usage by $1.5$--$3.2$ times when compared to unsecured scheme --- this is expected since vanilla execution does not provide any security guarantees (and there is no additional context switch overhead). We also note that \pnametee \textit{reduces CPU load by $30.48\%$--$47.32\%$} when compared to the fine-grain scheme;
this could be useful for many applications (say for battery operated systems to improve thermal efficiency).

\begin{center}
	\noindent\fbox{%
		\parbox{0.97\columnwidth}{%
			\em	 In comparsion with fine-grained checking, on average, \pnametee uses $30.48\%$--$47.32\%$ less CPU for its operations.
		}%
	}
\end{center}

Our experiments on the four real platforms conclude that \pnametee provides significant savings in task response time (\ie guarantees stability and ensures system safety) and resource usage but comes with a cost (\eg mean detection delay is at most $3$ jobs). Table \ref{tab:cs_res_summary} summarizes our findings (\ie delay in detection as well as reduction in execution time and CPU usage) for all four experimental platforms. As we see in our experiments, there is a trade-off between security and real-time requirements. For instance, the fine-grain checking scheme can detect the intrusions faster (\ie requires at most one additional instances) when compared to \pnametee. However it results in the controller tasks taking significantly longer time to finish (\ie can make the system unstable) and consume more resources (\eg CPU, battery, memory). By using the mechanisms proposed in \pnametee, \textit{designers can now measure such trade-offs}, evaluate the cost of integrating security and customize the number of security checks for a task that provides the best balance between real-time and security guarantees.

\begin{table}[!t]
\caption{Comparison with Fine-grain Checking: Summary of Findings}
\label{tab:cs_res_summary}
\centering
\begin{tabular}{P{1.95cm} ||  P{1.85cm} | P{1.40cm} | P{0.8cm} | P{0.80cm}}
\hline
\bfseries Platform & \multicolumn{4}{l}{\bfseries Performance Metrics: \pnametee vs Fine-grain}\\
\hline \hline
    & Execution Time Reduction &  CPU Usage Reduction &   \multicolumn{2}{P{2.1cm}}{Detection Delay (Task Instances)} \\  \cline{4-5}
  & (\%)  & (\%) & Mean & 99\textsuperscript{th}-p \\
\hline
Ground Rover  & $49.69$ & $37.58$ & $1$ & $3$ \\
Flight Controller  & $59.81$ & $47.32$ & $2$ & $8$ \\
Robotic Arm  & $42.80$ & $30.48$ & $3$ & $12$ \\
Syringe Pump & $56.81$ & $42.87$  & $2$ & $8$ \\
\hline
\end{tabular}
\end{table}

\begin{table}[!t]
\caption{Simulation Parameters}
\label{tab:ex_param_syn}
\centering
\begin{tabular}{p{5.0cm}||l}
\hline 
\bfseries Parameter & \bfseries Values\\
\hline\hline
Number of processor cores, $P$ & $4$ \\
Number of tasks, $M$ & $[12, 40]$ \\
Task periods, $T_i$ & $[10, 1000]$ ms \\
Number of actuation requests, $N_i$ & $\{ [3,5],~[8,10] \} $\\
Minimum requests verified per job, $N_i^{min}$ & $ \left\lceil 0.2N_i \right\rceil $\\
Verification overhead, $C_i^o$ & $10\%$ of $C_i$ \\
\hline
\end{tabular}
\end{table}

\subsection{Simulation-based Evaluation}
\label{sec:syn_experiments}

We also developed an open source simulator~\cite{rt_act_sec_game_repo} and conducted experiments with randomly generated workloads for a broader design-space exploration. Table~\ref{tab:ex_param_syn} lists the parameters used in our simulations.

\subsubsection{Workload Generation}

We considered $P = 4$ cores and each taskset instance contained $[3P, 10P]$ tasks. To generate systems with an even distribution
of tasks, we grouped the tasksets by base CPU utilization\footnote{In real-time terminology, the \textit{utilization} of a task is given by the ratio of its execution time to period~\cite{Liu_n_Layland1973}.} from $[(0.01+0.1 i)P, (0.1+0.1i)P]$ where $i \in \mathbb{Z}, 0 \leq i \leq 9$. Each utilization group contained $500$ tasksets (\ie a total $10 \times 500 = 5000$ tasksets were tested). We assumed that the tasks were partitioned using the first-fit strategy~\cite{parti_see}. We only considered the feasible tasksets (\ie the ones where response times are less than deadlines for all tasks) --- since tasksets that fail to meet this condition are trivially unschedulable. Task periods were generated according to a log-uniform distribution where each task had periods between $[10, 1000]$ ms. We assumed rate-monotonic priority ordering (\ie shorter period implies higher priorities)~\cite{Liu_n_Layland1973} since this is the standard techniques used in practical applications~\cite{mhasan_rtss16,mhasan_ecrts17,mhasan_date18,mhasan_date20}.
For a given number of tasks and total system utilization, the utilization of individual tasks were generated using Randfixedsum algorithm~\cite{randfixedsum}. 

We further assumed that for each task $\tau_i$, the overhead for checking each actuation command ($C_i^o$) is no more than $10\%$ of task execution time $C_i$ (\ie $C_i^o = 0.1C_i$). We considered two actuation command request scenarios: \ci medium ($N_i \in [3, 5],~\forall \tau_i$) and \cii high ($N_i \in [8, 10],~\forall \tau_i$).  We also assumed equal weights for all actuation commands and the minimum number of checking $N_i^{min}$ was at least $20\%$ of total number of requests (\ie  $N_i^{min} = \left\lceil 0.2N_i \right\rceil$).

\subsubsection{Results} \label{sec:syn_results}

\begin{figure}[!t]
\centering
\includegraphics[scale=0.50]{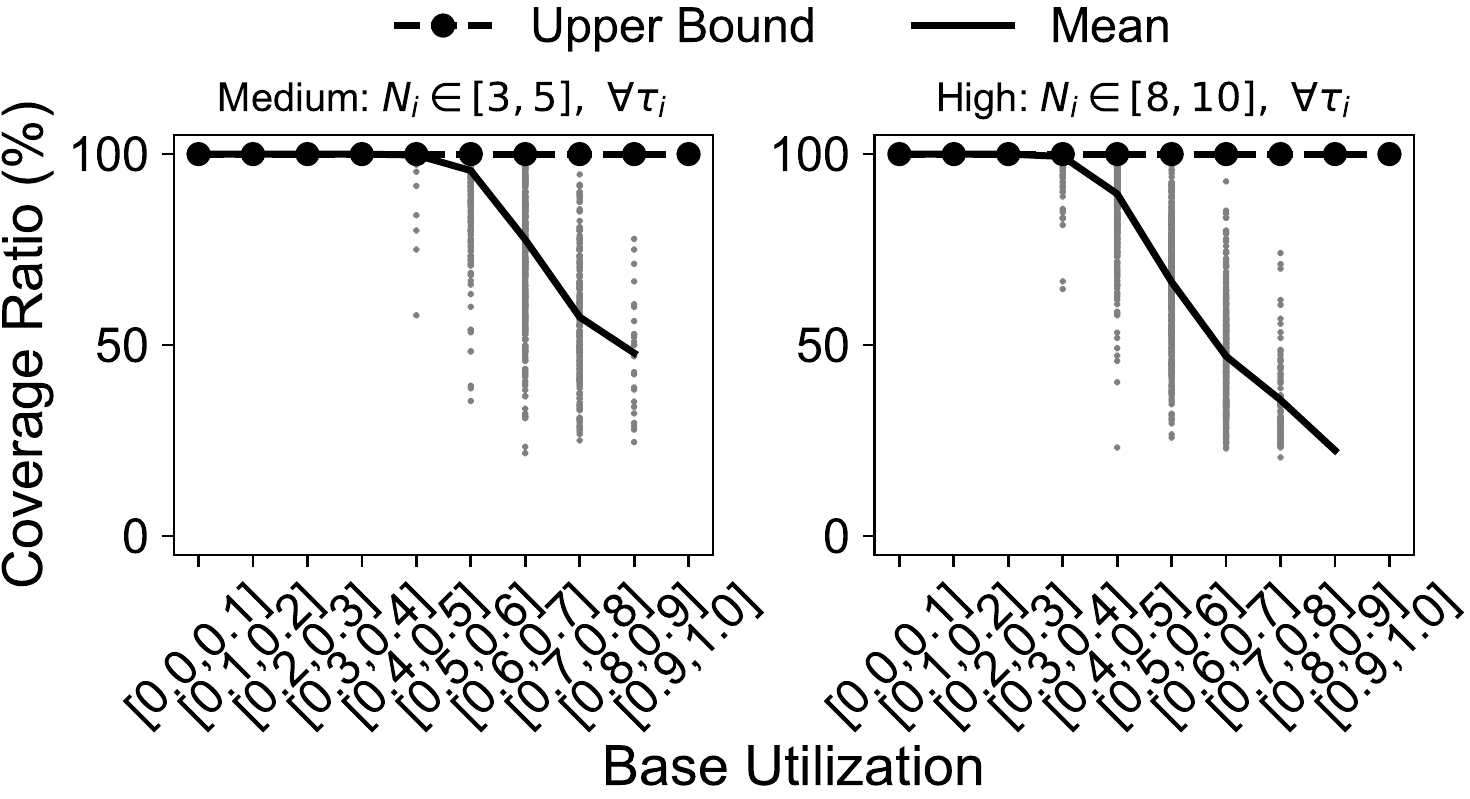}
\caption{Utilization vs coverage ratio. The upper bound represents fine grain checking where all the commands are verified. \pnametee can provide at least $50\%$ coverage per instance if the system utilization is no more than $70\%$. } 
\label{fig:covrat}
\end{figure}

In the following experiments (Fig.~\ref{fig:covrat}) we study how many commands can be checked in each of the task instances. For this, we introduce a metric called ``\textit{coverage ratio}'' ($CR$). The $CR$ metric shows us how many actuation commands (out of the total number of commands) we can check without violating timing constraints. We define  $CR$ as follows: $CR = \tfrac{1}{|\Gamma|}\sum_{\tau_i \in \Gamma} \tfrac{K_i}{N_i}$ where $\tfrac{1}{|\Gamma|}\sum_{\tau_i \in \Gamma} \tfrac{N_i^{min}}{N_i} \leq CR \leq 1$ $|\Gamma|$ is the total number of tasks and the parameter $K_i$ is obtained from Algorithm~\ref{alg:multicore_act_verif} (see Sec.~\ref{sec:cal_ki}). Notice that $CR=1$ (\ie upper bound) represents the fine-grain checking case since we verify all the commands. Let us now define base-utilization  of a taskset (\ie total utilization without any actuation checking) as $\tfrac{U}{P}$ where $U = \sum_{\tau_i \in \Gamma} \tfrac{C_i}{T_i}$, $C_i$ is the task execution time and $P_i$ is the period. The x-axis of Fig.~\ref{fig:covrat} shows the base-utilization and y-axis shows coverage ratio for both medium (left plot) and high (right plot) actuation scenarios. From our experiments we find that \textit{\pnametee provides a similar level of security when compared to fine-grain scheme} if the total utilization is no more than $60\%$ and $40\%$ for medium and high actuation request cases, respectively. We note that while fine-grain checking can provide better coverage, this upper bound is otherwise unattainable (specially for high utilization cases) since the all the tasks may not meet their timing requirements. This is seen in our additional experiments (Fig.~\ref{fig:sched}). In contrast, \pnametee can provide at least $50\%$ coverage even in high utilization cases (\eg at $\tfrac{U}{P} \leq 0.7$), when fine-grained checking fails since it makes the system infeasible.

\begin{center}
	\noindent\fbox{%
		\parbox{0.97\columnwidth}{%
			\em	 The performance of \pnametee is identical to the fine-grain scheme for low-to-medium system utilizations (\ie able to check all actuation commands). 
		}%
	}
\end{center}

\begin{figure}[!t]
\centering
\includegraphics[scale=0.45]{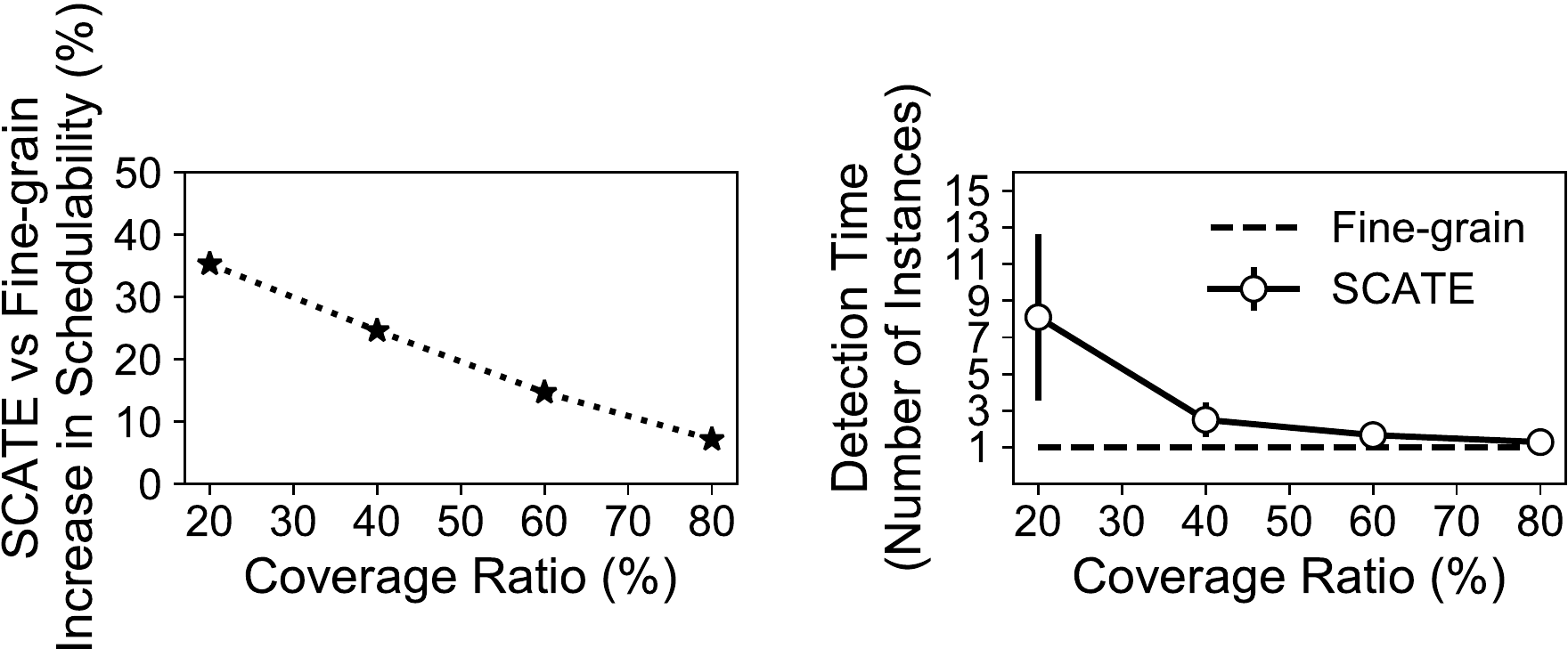}
\caption{Real-time and security trade-offs: low coverage ratio (\ie when fewer commands are checked in each task instance), while increasing the acceptance ratio (left plot), can lead to increase in detection time (right plot). We set the number of actuation commands at $N_i=5$.} 
\label{fig:d_time}
\end{figure}

In the last set of experiments (Fig.~\ref{fig:d_time}) we show the trade-off between real-time and security guarantees. For this, we use the ``schedulability'' metric --- a useful mathematical tool developed by the real-time community to analyze whether all activities of a  given system can meet their timing constraints even in the worst-case behavior of the system~\cite{mutiprocessor_survey}. A given taskset is considered as \textit{schedulable} if \textit{all} the tasks in the taskset meet their timing requirements (\ie response time is less than or equal to deadline). If the tasks are not schedulable, the system will be unsafe (and should not be deployed). The x-axis of Fig.~\ref{fig:d_time} shows the coverage ratio. The y-axis in the left figure shows the increase in schedulability in \pnametee when compared to the fine-grain scheme while the right figure shows the detection time (in terms of number of task instances) for both of the schemes. As we mentioned earlier, in the fine-grain scheme, the detection time for a known attack is upper bounded by the period of the task (\ie requires at most one additional instances when compared to the insecure base-case). As we see from the figure, there is a trade-off between real-time and security requirements: lower coverage ratio increases the schedulability (since there are lower checking overheads) but increases the detection times. This is because if coverage ratio is low, only a few commands are selected for checking during each instance and a vulnerable/compromised command will only be verified infrequently; thus resulting in longer detection times. We also carried out additional experiments (Appendix~\ref{sec:sched_plot}) to study the trade-offs of integrating TEE-based \pnametee mechanism in existing real-time platforms. 
Our results (see Fig.~\ref{fig:sched} in Appendix~\ref{sec:sched_plot}) show that there is a cost of integrating security (since it reduces the number of tasks that meets their timing requirements).

\begin{center}
	\noindent\fbox{%
		\parbox{0.97\columnwidth}{%
			\em	 If we perform fewer checks in each task instance, we can accommodate more tasks in the system (\ie higher acceptance ratio). However, this may result in delayed detections (\eg on average, requires eight additional task instances) since not all the actuation commands are frequently checked. 
		}%
	}
\end{center}

{\bf Summary.} Our experiments reveal interesting trade-offs between real-time and security requirements. Fine-grain checking --- while providing better security guarantees (\ie lower detection time) --- can negatively affect schedulability and, hence, safety and integrity of the system. \pnametee, in contrast, provides better schedulability guarantees (especially for high utilization cases) but may be result in slower detection times. By using our approach, designers of the systems can now customize their  platforms and selectively verify actuation commands based on application requirements. 
\section{Discussion} \label{sec:discussion}


 
 Our current implementation blocks malicious commands. Other strategies could involve the raising of alarms and/or sending out buffered (or even predetermined) alternate commands.  \pnametee transfers the control to the secure enclave for a (random) subset of actuation requests. Although this does not jeopardize the safety of the physical system (see more in the following paragraph), an adversary can still send spoofed signals in lieu of the unchecked commands (perhaps in a brute force manner). Alternative design choices (to further improve security and minimize overhead) could \ci pass of all the actuation requests through the trusted enclave
 	but checks only a few commands or \cii buffer multiple commands together and then transfer control \textit{once} to the enclave for checking.  We intend to incorporate these features in future work. 
 
While \pnametee adds delays in detection (\eg on average $1$--$3$ task instances when compared to the scheme that checks all the actuation commands), we can still retain the safety and normal operations of the plant due to physical inertia. For example, consider a simplified drone example. If an adversary sends false commands and turns off the propellers, the drone's altitude will not instantaneously drop to zero. Hence, although \pnametee may not detect the attack instantly, it can still block spoofed commands and prevent the drone from crashing. We present additional results and discuss this topic further in Appendix~\ref{sec:phys_inertia}.

In this paper we 
assume the existence of ``perfect'' checking modules provided by the system engineers (\ie an attack is always correctly detected by the \checkact function). Depending on the actual implementation, \checkact functions may result in false-positive/false-negative errors. Our model can also handle such cases by incorporating the detection inefficiency factors in the calculation of reward/cost metrics. For example, if the detection accuracy of \checkact  is $95\%$, one way to express reward and cost functions is as follows: $\lambda_{\rm{Imperfect}} = (1-.05)\lambda$ and $\zeta_{\rm{Imperfect}} = (1+.05)\zeta$, respectively.

 The game theory-based formulation used in our work can be extended to other real-time security use-cases. For instance, consider a distributed CPS where real-time nodes periodically exchange messages that need to be encrypted/authenticated (say to prevent man-in-the-middle attacks)~\cite{lin2009static,xie2007improving,lesi2017network}. 
While longer key sizes can provide better encryption, it requires more time to perform crypto operations (hence less number of messages can be secured and/or tasks can miss deadlines). By using game formulations similar to those in this paper, designers of systems can use different (perhaps smaller) key sizes for different messages (to reduce overhead) that provides maximal security from an external observer's point of view while guaranteeing timing requirements. 
%

Although we demonstrate our ideas on four realistic real-time platforms, the lack of real-time benchmarks is one of the major challenges in evaluating real-time cyber-physical security solutions. This is partly because of the diversity of such applications and software/hardware platforms as well as their hardware-dependent nature. In addition, critical cyber-physical applications are rarely open-sourced for safety/security/proprietary reasons. As a result, existing real-time security research is mainly evaluated by limited case-studies~\cite{mohan_s3a,securecore, securecore_memory,securecore_syscal,mhasan_resecure_iot,mhasan_rtss16,mhasan_ecrts17,mhasan_date18,mhasan_date20,mhasan_iotsnp19,lesi2017security,lesi2017network} and/or by using simulations~\cite{sg1,sg2,lin2009static,xie2007improving}.




\section{Related Work} \label{sec:related_work}

Enhancing security in time-critical cyber-physical systems is an active research area~\cite{mhasan_rtiot_sensors19,chai2019short}. 
Perhaps the closest line of work to ours is PROTC~\cite{protc_liu2017} where a monitor in the enclave
enforces secure access control policy (given by the control center) for
some peripherals of the drone and ensures
that only authorized applications can access
certain peripherals. Unlike our scheme, PROTC is limited for specific applications (\ie aerial robotic vehicles), requires a centralized control center to validate/enforce security policies and does not consider any load balancing issues. In other work~\cite{mhasan_ecrts17, mhasan_rtss16, mhasan_date18, mhasan_date20}, we proposed mechanisms to secure legacy RTS by integrating additional, independent, periodic monitoring tasks. They are pure software-based solutions and do not guarantee the integrity of security checks like \pnametee.  In contrast, in this paper consider a TEE-based RTS where security checks (\ie actuation command verification using trusted enclave) are interleaved within real-time tasks. Researchers also proposed anomaly detection approaches for robotic vehicles~\cite{guo2018roboads, choi2018detecting, fei2018cross}. These approaches do not leverage the capabilities of TEEs (\ie are vulnerable if the adversary can compromise the host OS) and do not consider real-time aspects. There exist various hardware/software-based mechanisms and architectural frameworks~\cite{mohan_s3a, securecore, securecore_memory, securecore_syscal, mhasan_resecure16,  mhasan_resecure_iot, slack_cornell}. 
However they
are not designed to protect against control-specific attacks and may not be suitable for systems developed with COTS components.

Researchers also use game-theoretical analysis for \ca general-purpose control systems~\cite{moothedath2020game,chen2016game}, \cb decision making problems~\cite{yang2019adaptive} and \cc preventing physical intrusions in CPS~\cite{rass2017physical}. These schemes are not designed to protect the systems against false actuation commands. In addition, they are not timing-aware (\ie real-time requirements are not considered). 
There also exist large number of research for generic cyber-physical/IoT-specific embedded systems as well as use of TrustZone to secure traditional embedded/mobile applications (too many to enumerate here, refer to  the related surveys
~\cite{cps_security_sruvey_iot,yang2017survey, ammar2018internet, trustzone_survey, trustzone_survey_2}) 
--- however the consideration of actuation-specific security scheduling and overload management aspects of real-time applications distinguish our work from other research. To the best of our knowledge, this is the first comprehensive work that introduces the notion of randomized coarse-grain checking for overloaded systems in order to validate actuation commands in a TEE-enabled real-time CPS. 
\section{Conclusion} \label{sec:conclusion}

In this paper we present a framework, \pnametee, to enhance the security and safety of the time-critical cyber-physical systems. We use a combination of trusted hardware and the intrinsic real-time nature of such systems and propose techniques to selectively verify a subset of commands that provides a trade-offs between real-time and security guarantees. We believe that our technique can be incorporated into multiple cyber-physical application domains such as avionics, automobiles, industrial control systems, medical devices, unmanned and autonomous vehicles.


\bibliographystyle{IEEEtran}
\bibliography{IEEEabrv,references_short}

\begin{thebibliography}{100}
\providecommand{\url}[1]{#1}
\csname url@samestyle\endcsname
\providecommand{\newblock}{\relax}
\providecommand{\bibinfo}[2]{#2}
\providecommand{\BIBentrySTDinterwordspacing}{\spaceskip=0pt\relax}
\providecommand{\BIBentryALTinterwordstretchfactor}{4}
\providecommand{\BIBentryALTinterwordspacing}{\spaceskip=\fontdimen2\font plus
\BIBentryALTinterwordstretchfactor\fontdimen3\font minus
  \fontdimen4\font\relax}
\providecommand{\BIBforeignlanguage}[2]{{%
\expandafter\ifx\csname l@#1\endcsname\relax
\typeout{** WARNING: IEEEtran.bst: No hyphenation pattern has been}%
\typeout{** loaded for the language `#1'. Using the pattern for}%
\typeout{** the default language instead.}%
\else
\language=\csname l@#1\endcsname
\fi
#2}}
\providecommand{\BIBdecl}{\relax}
\BIBdecl

\bibitem{castelli2019development}
K.~Castelli, A.~M.~A. Zaki, and H.~Giberti, ``Development of a practical tool
  for designing multi-robot systems in pick-and-place applications,''
  \emph{MDPI Robotics}, vol.~8, no.~3, 2019.

\bibitem{mhasan_rtiot_sensors19}
C.-Y. Chen, M.~Hasan, and S.~Mohan, ``Securing real-time
  {Internet}-of-things,'' \emph{Sensors}, vol.~18, no.~12, 2018.

\bibitem{otuoze2018smart}
A.~O. Otuoze, M.~W. Mustafa, and R.~M. Larik, ``Smart grids security
  challenges: Classification by sources of threats,'' \emph{Elsevier JESIT},
  vol.~5, no.~3, pp. 468--483, 2018.

\bibitem{zheng2018smart}
P.~Zheng, Z.~Sang, R.~Y. Zhong, Y.~Liu, C.~Liu, K.~Mubarok, S.~Yu, X.~Xu
  \emph{et~al.}, ``Smart manufacturing systems for {Industry 4.0}: Conceptual
  framework, scenarios, and future perspectives,'' \emph{Springer Front. of
  Mech. Eng.}, vol.~13, no.~2, pp. 137--150, 2018.

\bibitem{azgomi2018brief}
H.~F. Azgomi and M.~Jamshidi, ``A brief survey on smart community and smart
  transportation,'' in \emph{IEEE ICTAI}, 2018, pp. 932--939.

\bibitem{securecore}
M.-K. Yoon, S.~Mohan, J.~Choi, J.-E. Kim, and L.~Sha, ``{SecureCore}: A
  multicore-based intrusion detection architecture for real-time embedded
  systems,'' in \emph{IEEE RTAS}, 2013, pp. 21--32.

\bibitem{mhasan_resecure_iot}
F.~Abdi, C.-Y. Chen, M.~Hasan, S.~Liu, S.~Mohan, and M.~Caccamo, ``Preserving
  physical safety under cyber attacks,'' \emph{IEEE IoT J.}, vol.~6, no.~4, pp.
  6285--6300, 2018.

\bibitem{mohan_s3a}
S.~Mohan, S.~Bak, E.~Betti, H.~Yun, L.~Sha, and M.~Caccamo, ``{S3A}: Secure
  system simplex architecture for enhanced security and robustness of
  cyber-physical systems,'' in \emph{ACM HiCoNS}.\hskip 1em plus 0.5em minus
  0.4em\relax ACM, 2013, pp. 65--74.

\bibitem{stuxnet}
N.~Falliere, L.~O. Murchu, and E.~Chien, ``W32. stuxnet dossier,'' \emph{White
  paper, Symantec Corp}, vol.~5, p.~6, 2011.

\bibitem{security_medical}
S.~S. Clark and K.~Fu, ``Recent results in computer security for medical
  devices,'' in \emph{MobiHealth}, 2011, pp. 111--118.

\bibitem{checkoway2011comprehensive}
S.~Checkoway, D.~McCoy, B.~Kantor, D.~Anderson, H.~Shacham, S.~Savage,
  K.~Koscher, A.~Czeskis, F.~Roesner, T.~Kohno \emph{et~al.}, ``Comprehensive
  experimental analyses of automotive attack surfaces,'' in \emph{USENIX Sec.
  Symp.}, 2011.

\bibitem{ddos_iot_camera}
J.~Westling, ``Future of the {Internet} of things in mission critical
  applications,'' 2016.

\bibitem{simmon2013vision}
E.~Simmon, K.-S. Kim, E.~Subrahmanian, R.~Lee, F.~De~Vaulx, Y.~Murakami,
  K.~Zettsu, and R.~D. Sriram, \emph{A vision of cyber-physical cloud computing
  for smart networked systems}.\hskip 1em plus 0.5em minus 0.4em\relax US Dept.
  of Commerce, NIST, 2013.

\bibitem{trustzone_survey}
S.~Pinto and N.~Santos, ``Demystifying {ARM TrustZone}: A comprehensive
  survey,'' \emph{ACM CSUR}, vol.~51, no.~6, p. 130, 2019.

\bibitem{intel_sgx}
V.~Costan and S.~Devadas, ``{Intel SGX Explained},'' \emph{IACR Crypt. ePrint
  Arch.}, no. 086, pp. 1--118, 2016.

\bibitem{harsanyi1972generalized}
J.~C. Harsanyi and R.~Selten, ``A generalized nash solution for two-person
  bargaining games with incomplete information,'' \emph{INFORMS Man. Sci.},
  vol.~18, no. 5-part-2, pp. 80--106, 1972.

\bibitem{conitzer2006computing}
V.~Conitzer and T.~Sandholm, ``Computing the optimal strategy to commit to,''
  in \emph{ACM EC}, 2006, pp. 82--90.

\bibitem{paruchuri2007efficient}
P.~Paruchuri, J.~P. Pearce, M.~Tambe, F.~Ordonez, and S.~Kraus, ``An efficient
  heuristic approach for security against multiple adversaries,'' in
  \emph{IFAAMAS AAMAS}, 2007, pp. 1--8.

\bibitem{rpi3}
``{Raspberry Pi},'' \url{https://tinyurl.com/rpi3modelb}.

\bibitem{rt_act_sec_game_repo}
``\pnametee implementation,''
  \url{https://github.com/mnwrhsn/scate_implementation}.

\bibitem{guo2018roboads}
P.~Guo, H.~Kim, N.~Virani, J.~Xu, M.~Zhu, and P.~Liu, ``{RoboADS: Anomaly
  detection against sensor and actuator misbehaviors in mobile robots},'' in
  \emph{IEEE/IFIP DSN}, 2018, pp. 574--585.

\bibitem{gpg2_lf}
``{Dexter Industries Sensors},'' \url{https://github.com/DexterInd/DI_Sensors}.

\bibitem{i2c}
\BIBentryALTinterwordspacing
``{I}\textsuperscript{2}{C} manual,'' Philips Semiconductors, 2003. [Online].
  Available: \url{https://tinyurl.com/i2c-manual}
\BIBentrySTDinterwordspacing

\bibitem{mahfouzi2019butterfly}
R.~Mahfouzi, A.~Aminifar, S.~Samii, M.~Payer, P.~Eles, and Z.~Peng, ``Butterfly
  attack: Adversarial manipulation of temporal properties of cyber-physical
  systems,'' in \emph{IEEE RTSS}, 2019, pp. 93--106.

\bibitem{mhasan_iotsnp19}
M.~Hasan and S.~Mohan, ``Protecting actuators in safety critical {IoT} systems
  from control spoofing attacks,'' in \emph{ACM IoT S\&P}, 2019, pp. 8--14.

\bibitem{cy_scheduleak_rtas19}
C.-Y. Chen, S.~Mohan, R.~Pellizzoni, R.~B. Bobba, and N.~Kiyavash, ``A novel
  side-channel in real-time schedulers,'' in \emph{IEEE RTAS}, 2019, pp.
  90--102.

\bibitem{liu2019leaking}
S.~Liu, N.~Guan, D.~Ji, W.~Liu, X.~Liu, and W.~Yi, ``Leaking your engine speed
  by spectrum analysis of real-time scheduling sequences,'' \emph{J. of Sys.
  Arch.}, vol.~97, pp. 455--466, 2019.

\bibitem{dos_heechal_rtas19}
M.~Bechtel and H.~Yun, ``Denial-of-service attacks on shared cache in
  multicore: Analysis and prevention,'' in \emph{IEEE RTAS}, 2019, pp.
  357--367.

\bibitem{lesi2017network}
V.~Lesi, I.~Jovanov, and M.~Pajic, ``Network scheduling for secure
  cyber-physical systems,'' in \emph{IEEE RTSS}, 2017, pp. 45--55.

\bibitem{lesi2017security}
V.~\vspace{0mm}Lesi, I.~Jovanov, and M.~Pajic, ``Security-aware scheduling of
  embedded control tasks,'' \emph{ACM TECS}, vol.~16, pp. 188:1--188:21, 2017.

\bibitem{loi2017systematically}
F.~Loi, A.~Sivanathan, H.~H. Gharakheili, A.~Radford, and V.~Sivaraman,
  ``{Systematically evaluating security and privacy for consumer IoT
  devices},'' in \emph{ACM IoTS\&P}, 2017, pp. 1--6.

\bibitem{sg2}
R.~Pellizzoni, N.~Paryab, M.-K. Yoon, S.~Bak, S.~Mohan, and R.~B. Bobba, ``A
  generalized model for preventing information leakage in hard real-time
  systems,'' in \emph{IEEE RTAS}, 2015, pp. 271--282.

\bibitem{securecore_syscal}
M.-K. Yoon, S.~Mohan, J.~Choi, M.~Christodorescu, and L.~Sha, ``Learning
  execution contexts from system call distribution for anomaly detection in
  smart embedded system,'' in \emph{ACM/IEEE IoTDI}, 2017, pp. 191--196.

\bibitem{kim2018securing}
C.~H. Kim, T.~Kim, H.~Choi, Z.~Gu, B.~Lee, X.~Zhang, and D.~Xu, ``Securing
  real-time microcontroller systems through customized memory view switching.''
  in \emph{NDSS}, 2018.

\bibitem{choi2018detecting}
H.~Choi, W.-C. Lee, Y.~Aafer, F.~Fei, Z.~Tu, X.~Zhang, D.~Xu, and X.~Xinyan,
  ``Detecting attacks against robotic vehicles: A control invariant approach,''
  in \emph{ACM CCS}, 2018, pp. 801--816.

\bibitem{hussain2006vehicle}
A.~Hussain, M.~Hannan, A.~Mohamed, H.~Sanusi, and A.~Ariffin, ``Vehicle crash
  analysis for airbag deployment decision,'' \emph{Int. J. of Auto. Tech.},
  vol.~7, no.~2, pp. 179--185, 2006.

\bibitem{algorithmic_game_theory}
T.~Roughgarden, ``Algorithmic game theory,'' \emph{Comm. of the ACM}, vol.~53,
  no.~7, pp. 78--86, 2010.

\bibitem{korilis1997achieving}
Y.~A. Korilis, A.~A. Lazar, and A.~Orda, ``Achieving network optima using
  {Stackelberg} routing strategies,'' \emph{IEEE/ACM TON}, vol.~5, no.~1, pp.
  161--173, 1997.

\bibitem{cardinal2005pricing}
J.~Cardinal, M.~Labb{\'e}, S.~Langerman, and B.~Palop, ``Pricing of geometric
  transportation networks,'' in \emph{CCCG}, 2005, pp. 92--96.

\bibitem{moothedath2020game}
S.~Moothedath, D.~Sahabandu, J.~Allen, A.~Clark, L.~Bushnell, W.~Lee, and
  R.~Poovendran, ``A game-theoretic approach for dynamic information flow
  tracking to detect multi-stage advanced persistent threats,'' \emph{IEEE
  TACON}, 2020.

\bibitem{yang2019adaptive}
G.~Yang, R.~Poovendran, and J.~P. Hespanha, ``Adaptive learning in two-player
  stackelberg games with continuous action sets,'' in \emph{IEEE CDC}, 2019,
  pp. 6905--6911.

\bibitem{chen2016game}
J.~Chen and Q.~Zhu, ``A game-theoretic framework for resilient and distributed
  generation control of renewable energies in microgrids,'' \emph{IEEE Trans.
  on Smart Grid}, vol.~8, no.~1, pp. 285--295, 2016.

\bibitem{rass2017physical}
S.~Rass, A.~Alshawish, M.~A. Abid, S.~Schauer, Q.~Zhu, and H.~De~Meer,
  ``Physical intrusion games—optimizing surveillance by simulation and game
  theory,'' \emph{IEEE Access}, vol.~5, pp. 8394--8407, 2017.

\bibitem{yu2015handling}
T.~Yu, V.~Sekar, S.~Seshan, Y.~Agarwal, and C.~Xu, ``Handling a trillion
  (unfixable) flaws on a billion devices: Rethinking network security for the
  {I}nternet-of-things,'' in \emph{ACM HotNets}, 2015, pp. 1--7.

\bibitem{adepu2017design}
S.~Adepu and A.~Mathur, ``From design to invariants: Detecting attacks on cyber
  physical systems,'' in \emph{IEEE QRS-C}, 2017, pp. 533--540.

\bibitem{berthier2011specification}
R.~Berthier and W.~H. Sanders, ``Specification-based intrusion detection for
  advanced metering infrastructures,'' in \emph{IEEE PRDC}.\hskip 1em plus
  0.5em minus 0.4em\relax IEEE, 2011, pp. 184--193.

\bibitem{anomaly_detection_survey}
V.~Chandola, A.~Banerjee, and V.~Kumar, ``Anomaly detection: A survey,''
  \emph{ACM CSUR}, vol.~41, no.~3, p.~15, 2009.

\bibitem{mukherjee2019optimized}
A.~Mukherjee, T.~Mishra, T.~Chantem, N.~Fisher, and R.~Gerdes, ``Optimized
  trusted execution for hard real-time applications on cots processors,'' in
  \emph{ACM RTNS}, 2019, pp. 50--60.

\bibitem{amacher2019performance}
J.~Amacher and V.~Schiavoni, ``On the performance of arm trustzone,'' in
  \emph{IFIP DAIS}, 2019, pp. 133--151.

\bibitem{liu2018rt}
Y.~Liu, K.~An, and E.~Tilevich, ``{RT-trust: Automated refactoring for trusted
  execution under real-time constraints},'' in \emph{ACM GPCE}, 2018, pp.
  175--187.

\bibitem{optee}
``{Open Portable Trusted Execution Environment},''
  \url{https://www.op-tee.org/}.

\bibitem{arm_fvp}
``{ARM Fixed Virtual Platforms},'' 
  \url{https://developer.arm.com/tools-and-software/simulation-models/fixed-virtual-platforms}.

\bibitem{free_rtos}
``{FreeRTOS},'' \url{http://www.freertos.org}.

\bibitem{res_time_rts}
N.~Audsley, A.~Burns, M.~Richardson, K.~Tindell, and A.~J. Wellings, ``Applying
  new scheduling theory to static priority pre-emptive scheduling,'' \emph{SE
  Journal}, vol.~8, no.~5, pp. 284--292, 1993.

\bibitem{cheng2017orpheus}
L.~Cheng, K.~Tian, and D.~D. Yao, ``Orpheus: Enforcing cyber-physical execution
  semantics to defend against data-oriented attacks,'' in \emph{ACM ACSAC},
  2017, pp. 315--326.

\bibitem{virtsense_liu2018}
R.~Liu and M.~Srivastava, ``{VirtSense: Virtualize Sensing through ARM
  TrustZone on Internet-of-Things},'' in \emph{ACM SysTEX}, 2018, pp. 2--7.

\bibitem{protc_liu2017}
R.~\vspace*{0em}Liu and M.~Srivastava, ``{PROTC: PROTeCting drone's peripherals
  through ARM trustzone},'' in \emph{ACM DroNet}, 2017, pp. 1--6.

\bibitem{liu2018alidrone}
T.~Liu, A.~Hojjati, A.~Bates, and K.~Nahrstedt, ``Alidrone: Enabling
  trustworthy proof-of-alibi for commercial drone compliance,'' in \emph{IEEE
  ICDCS}, 2018, pp. 841--852.

\bibitem{adafruit_motor_hat}
``{Adafruit motor shield},''
  \url{https://learn.adafruit.com/adafruit-motor-shield}.

\bibitem{adafruit_motor_hat_code}
``{Adafruit driver},''
  \url{https://github.com/threebrooks/AdafruitStepperMotorHAT_CPP}.

\bibitem{pca9685}
``{PCA9685 I2C PWM driver},'' \url{https://github.com/TeraHz/PCA9685}.

\bibitem{globalplatform_tee_api}
``{TEE API},''
  \url{https://globalplatform.org/specs-library/tee-client-api-specification/}.

\bibitem{python_mip}
``{The Python-MIP package},'' \url{https://www.python-mip.com/}.

\bibitem{cbc_solver}
K.~Martin, ``{Tutorial: COIN-OR: Software for the OR community},''
  \emph{INFORMS Interfaces}, vol.~40, no.~6, pp. 465--476, 2010.

\bibitem{lipowski2012roulette}
A.~Lipowski and D.~Lipowska, ``Roulette-wheel selection via stochastic
  acceptance,'' \emph{Elsevier Physica A}, vol. 391, no.~6, pp. 2193--2196,
  2012.

\bibitem{Z1FFER}
``{Z1FFER open source hardware random number generator},''
  \par\url{http://www.openrandom.org}.

\bibitem{OneRNG}
``Open hardware random number generator,'' \url{https://onerng.info}.

\bibitem{artinali}
M.~R. Aliabadi, A.~A. Kamath, J.~Gascon-Samson, and K.~Pattabiraman,
  ``Artinali: dynamic invariant detection for cyber-physical system security,''
  in \emph{ACM ESEC/FSE}, 2017, pp. 349--361.

\bibitem{tabrizi2015flexible}
F.~M. Tabrizi and K.~Pattabiraman, ``Flexible intrusion detection systems for
  memory-constrained embedded systems,'' in \emph{IEEE EDCC}, 2015, pp. 1--12.

\bibitem{mhasan_resecure_iccps}
F.~Abdi, C.-Y. Chen, M.~Hasan, S.~Liu, S.~Mohan, and M.~Caccamo, ``Guaranteed
  physical security with restart-based design for cyber-physical systems,'' in
  \emph{ACM/IEEE ICCPS}, 2018, pp. 10--21.

\bibitem{rpi_rover_using_motor_hat}
``{Raspberry Pi rover},'' \url{https://github.com/Veilkrand/simplePiRover}.

\bibitem{gpg2}
``{GoPiGo},'' \url{https://github.com/DexterInd/GoPiGo}.

\bibitem{drone_flight_controller}
``{Drone controller},''
  \url{https://github.com/lobodol/drone-flight-controller}.

\bibitem{aastrom2004revisiting}
K.~J. {\AA}str{\"o}m and T.~H{\"a}gglund, ``{Revisiting the Ziegler--Nichols
  step response method for PID control},'' \emph{Elsevier J. of proc. con.},
  vol.~14, no.~6, pp. 635--650, 2004.

\bibitem{robot_arm_src}
``{Robot arm control},'' \url{https://github.com/tutRPi/6DOF-Robot-Arm}.

\bibitem{c-flat_impl}
``{C-FLAT implementation},''
  \url{https://github.com/control-flow-attestation/c-flat}.

\bibitem{abera2016c}
T.~Abera, N.~Asokan, L.~Davi, J.-E. Ekberg, T.~Nyman, A.~Paverd, A.-R. Sadeghi,
  and G.~Tsudik, ``{C-FLAT}: control-flow attestation for embedded systems
  software,'' in \emph{ACM CCS}, 2016, pp. 743--754.

\bibitem{xie2007improving}
T.~Xie and X.~Qin, ``Improving security for periodic tasks in embedded systems
  through scheduling,'' \emph{ACM TECS}, vol.~6, no.~3, p.~20, 2007.

\bibitem{sg1}
S.~Mohan, M.-K. Yoon, R.~Pellizzoni, and R.~B. Bobba, ``Real-time systems
  security through scheduler constraints,'' in \emph{Euromicro ECRTS}, 2014,
  pp. 129--140.

\bibitem{delay_period}
E.~Bini and A.~Cervin, ``Delay-aware period assignment in control systems,'' in
  \emph{IEEE RTSS}, 2008, pp. 291--300.

\bibitem{Liu_n_Layland1973}
C.~L. Liu and J.~W. Layland, ``Scheduling algorithms for multiprogramming in a
  hard-real-time environment,'' \emph{JACM}, vol.~20, no.~1, pp. 46--61, 1973.

\bibitem{parti_see}
J.~Chen, ``Partitioned multiprocessor fixed-priority scheduling of sporadic
  real-time tasks,'' in \emph{Euromicro ECRTS}, 2016, pp. 251--261.

\bibitem{mhasan_rtss16}
M.~Hasan, S.~Mohan, R.~B. Bobba, and R.~Pellizzoni, ``Exploring opportunistic
  execution for integrating security into legacy hard real-time systems,'' in
  \emph{IEEE RTSS}, 2016, pp. 123--134.

\bibitem{mhasan_ecrts17}
M.~\vspace{0mm}Hasan, S.~Mohan, R.~Pellizzoni, and R.~B. Bobba, ``Contego: An
  adaptive framework for integrating security tasks in real-time systems,'' in
  \emph{Euromicro ECRTS}, 2017, pp. 23:1--23:22.

\bibitem{mhasan_date18}
M.~Hasan, S.~Mohan, R.~Pellizzoni, and R.~B. Bobba, ``A design-space
  exploration for allocating security tasks in multicore real-time systems,''
  in \emph{DATE}, 2018, pp. 225--230.

\bibitem{mhasan_date20}
M.~\vspace*{0em}Hasan, S.~Mohan, R.~Pellizzoni, and R.~B. Bobba, ``Period
  adaptation for continuous security monitoring in multicore systems,'' in
  \emph{DATE}, 2020.

\bibitem{randfixedsum}
P.~Emberson, R.~Stafford, and R.~I. Davis, ``Techniques for the synthesis of
  multiprocessor tasksets,'' in \emph{WATERS}, 2010, pp. 6--11.

\bibitem{mutiprocessor_survey}
R.~I. Davis and A.~Burns, ``A survey of hard real-time scheduling for
  multiprocessor systems,'' \emph{ACM CSUR}, vol.~43, no.~4, pp. 35:1--35:44,
  2011.

\bibitem{lin2009static}
M.~Lin, L.~Xu, L.~T. Yang, X.~Qin, N.~Zheng, Z.~Wu, and M.~Qiu, ``Static
  security optimization for real-time systems,'' \emph{IEEE Trans. on Indust.
  Info.}, vol.~5, no.~1, pp. 22--37, 2009.

\bibitem{securecore_memory}
M.-K. Yoon, S.~Mohan, J.~Choi, and L.~Sha, ``Memory heat map: anomaly detection
  in real-time embedded systems using memory behavior,'' in \emph{ACM/EDAC/IEEE
  DAC}, 2015, pp. 1--6.

\bibitem{chai2019short}
H.~Chai, G.~Zhang, J.~Zhou, J.~Sun, L.~Huang, and T.~Wang, ``A short review of
  security-aware techniques in real-time embedded systems,'' \emph{J. of Cir.,
  Sys. and Comp.}, vol.~28, no.~02, 2019.

\bibitem{fei2018cross}
F.~Fei, Z.~Tu, R.~Yu, T.~Kim, X.~Zhang, D.~Xu, and X.~Deng, ``Cross-layer
  retrofitting of {UAVs} against cyber-physical attacks,'' in \emph{IEEE ICRA},
  2018, pp. 550--557.

\bibitem{mhasan_resecure16}
F.~Abdi, M.~Hasan, S.~Mohan, D.~Agarwal, and M.~Caccamo, ``{ReSecure}: A
  restart-based security protocol for tightly actuated hard real-time
  systems,'' in \emph{IEEE CERTS}, 2016, pp. 47--54.

\bibitem{slack_cornell}
D.~Lo, M.~Ismail, T.~Chen, and G.~E. Suh, ``Slack-aware opportunistic
  monitoring for real-time systems,'' in \emph{IEEE RTAS}, 2014, pp. 203--214.

\bibitem{cps_security_sruvey_iot}
A.~{Humayed}, J.~{Lin}, F.~{Li}, and B.~{Luo}, ``Cyber-physical systems
  security -- {A} survey,'' \emph{IEEE IoT J.}, vol.~4, no.~6, pp. 1802--1831,
  2017.

\bibitem{yang2017survey}
Y.~Yang, L.~Wu, G.~Yin, L.~Li, and H.~Zhao, ``{A survey on security and privacy
  issues in Internet-of-Things},'' \emph{IEEE IoT J.}, vol.~4, no.~5, pp.
  1250--1258, 2017.

\bibitem{ammar2018internet}
M.~Ammar, G.~Russello, and B.~Crispo, ``{Internet of Things: A survey on the
  security of IoT frameworks},'' \emph{Elsevier J. of Inf. Sec. \& App.},
  vol.~38, pp. 8--27, 2018.

\bibitem{trustzone_survey_2}
W.~Li, H.~Chen, and H.~Chen, ``{Research on ARM TrustZone},'' \emph{ACM
  GetMobile}, vol.~22, no.~3, pp. 17--22, 2019.

\bibitem{qnx}
F.~Kolnick, ``The {QNX} 4 real-time operating system,'' \emph{Basis Comp. Sys.
  Inc.}, 1998.

\bibitem{okl4}
G.~Heiser and B.~Leslie, ``The {OKL4} microvisor: Convergence point of
  microkernels and hypervisors,'' in \emph{ACM APSys}.\hskip 1em plus 0.5em
  minus 0.4em\relax ACM, 2010, pp. 19--24.

\bibitem{rt_patch}
L.~Fu and R.~Schwebel, ``Real-time {Linux} wiki,''
  \url{https://rt.wiki.kernel.org/index.php/rt_preempt_howto}, [Online].

\bibitem{wcrt_survey}
R.~Wilhelm, J.~Engblom, A.~Ermedahl, N.~Holsti, S.~Thesing, D.~Whalley,
  G.~Bernat, C.~Ferdinand, R.~Heckmann, T.~Mitra \emph{et~al.}, ``The
  worst-case execution-time problem—overview of methods and survey of
  tools,'' \emph{ACM TECS}, vol.~7, no.~3, p.~36, 2008.

\bibitem{isovic2001handling}
D.~Isovic, ``Handling sporadic tasks in real-time systems: Combined offline and
  online approach,'' Tech. Rep., June 2001.

\bibitem{cvxopt}
\BIBentryALTinterwordspacing
L.~Vandenberghe, ``{The CVXOPT linear and quadratic cone program solvers},''
  2010. [Online]. Available: \url{http://cvxopt.
  org/documentation/coneprog.pdf}
\BIBentrySTDinterwordspacing

\bibitem{luukkonen2011modelling}
T.~Luukkonen, ``Modelling and control of quadcopter,'' School of Science, Aalto
  University, Tech. Rep., 2011.

\end{thebibliography}


\appendix

\subsection{System Model: Real-Time Task Model and Scheduling} \label{sec:sys_mod_formal}

We consider a system that consists of $M$ fixed-priority real-time tasks $\Gamma = \{ \tau_i, \cdots, \tau_M \}$ running on $P$ identical processor cores $\Pi = \{\pi_1, \cdots, \pi_P\}$. In this work, we consider partitioned fixed-priority preemptive scheduling \cite{mutiprocessor_survey} (a widely supported approach in many commercial and open-source real-time operating systems such as QNX~\cite{qnx}, OKL4~\cite{okl4}, real-time Linux~\cite{rt_patch}, \etc) where tasks are statically assigned to the processor cores using a predefined partitioning scheme. The set of tasks running on a given core $\pi_p$ is denoted by $\Gamma_p$ and $\Gamma = \cup_{\pi_p \in \Pi} \{ \Gamma_p \}$. Each task $\tau_i$ is represented by the following tuple: $$(C_i, T_i, D_i, N_i, N_i^{min}, W_i).$$ 
Each of the above variables represents the following: 

\begin{itemize}
	\item $C_i$ is the constant, upper bound on the computation time, called worst-case execution time (WCET)~\cite{wcrt_survey};
	\item $T_i$ is the minimum inter-arrival time (period), \ie consecutive jobs of $\tau_i$ should be temporally separated by at least $T_i$ time units;
	\item $D_i$ (usually less than or equal to $T_i)$ is the timing constraint (deadline);
	\item $N_i$ is the number of actuation requests that $\tau_i$ sends out;
	\item $N_i^{min} \leq N_i$ is a QoS parameter that denotes the minimum number of actuation commands that must be checked; and
	\item $W_i = \{\omega_i^1, \cdots, \omega_i^{N_i} \}$ is a designer-provided weight vector where the weight $\omega_i^j$ represents the importance of $j$-th actuation command. 
\end{itemize}
 As shown in Sec.~\ref{sec:overload_control}, parameters $N_i^{min}$ and $W_i$ help the designers to determine the subset of commands to be selected in each iteration of the task for checking when not all $N_i$ commands can be checked due to timing constraints. While we present the above task model for ease of presentation, we note that not all the real-time tasks in a given system will send out actuation requests. For such tasks $\tau_{i^\prime}$ setting $N_{i^\prime} = N_{i^\prime}^{min} = 0$ and ignoring the variable $W_{i^\prime}$ will retain the consistency of the model. 


We consider a discrete time model~\cite{isovic2001handling} where the system and task parameters are multiples of a fixed time unit, \ie an interval starting from time point $t_1$ and ending at time point $t_2$ that has a length of $t_2-t_1$ by $[t_1,t_2)$ or $[t_1,t_2-1]$. We also assume that the non-secure system (\ie when there is no checking) is ``schedulable'', that is, for each task $\tau_i \in \Gamma$, the response time (time between completion and arrival) is less than the deadline of the task. Table~\ref{tab:rt_glossary} summarizes the real-time terminology used in the paper.

\begin{table}[!t]
	\caption{Real-Time Terminology}
	\label{tab:rt_glossary}
	\centering
	\begin{tabular}{P{4.30cm} P{11.00cm}}
		\hline
		\bfseries Term & \bfseries Interpretation\\
		\hline
		\hline 
		Deadline & Real-time constraint of a task. Tasks must finish their computation before deadlines to ensure the safe operation of the system \\ \hline 
		Period  & Inter-arrival time of a task. Each task instance (also called ``job'') periodically performs desired computation \\ \hline
		Schedulability & Mathematical tool to determine whether all tasks can meet their timing constraints (deadlines) even in the worst-case behavior of the system \\ \hline 
		Utilization & Ratio between task execution time to its period \\ \hline
		Worst-case execution time (WCET) & An upper bound of the task execution time, determined at the system design phase \\ \hline 
		Worst-case response time (WCRT) & Time between task arrival to completion. WCRT must be less than the deadline of the tasks to ensure schedulability \\
		\hline 
	\end{tabular}
\end{table}

\begin{table*}
\colorlet{mhgray}{lightgray!45}
\scriptsize
\centering
\caption{Actuation Command Checking for Various Cyber-Physical Platforms\textsuperscript{*}}
\label{tab:rt_act_exp}
\begin{tabular}{P{3.0cm} | P{3.0cm} |  P{1.2cm} |  P{8.5cm}} \hline 
\bf Platform & \bf Application & \bf Actuators & \bf Possible Checking Conditions \\
\hline \hline
\cellcolor{mhgray}Robotic vehicle \par (ground, aerial) & \cellcolor{mhgray}Surveillance, agriculture, manufacturing & \cellcolor{mhgray}Servo, motor & \cellcolor{mhgray}\ca Check if the robot is following the mission; \cb allow only predefined number of actuation commands per period \\ \hline
\cellcolor{mhgray}Robotic arm & \cellcolor{mhgray}Manufacturing & \cellcolor{mhgray}Servo,  buzzer & \cellcolor{mhgray}\ca Check the servo pulse sequences matches with the desired (design-time) sequence; \cb do not raise alarm if the pulse sequence is normal \\ \hline
\cellcolor{mhgray}Infusion/Syringe pump & \cellcolor{mhgray}Health-care & \cellcolor{mhgray}Motor, display & \cellcolor{mhgray}\ca Drive the motor only to allowable positions/rates \cb display only the amount of fluid infused (\eg obtained from motor encoders) \\ \hline
Water/air monitoring system & Home/industrial automation & Buzzer, display & \ca Send high pulse to buzzer only if water-level is high/air quality abnormal/detect smoke; \cb do not display alert if the system state is normal \\ \hline
Surveillance system & Home/industrial automation  & Servo, buzzer & \ca Trigger alarm only if there is an impact/object detected in camera; \cb rotate camera (using servos) only within allowable pan/tilt angle \\
\hline
\end{tabular}
\vspace*{-0.3em}
\begin{flushleft}
\textsuperscript{*}Platforms listed in shaded rows are implemented and evaluated in this work. Other examples are presented here to illustrate applicability of our ideas for multiple use-cases. 
\end{flushleft}
\end{table*}

\subsection{Examples of Checking Actuation Commands for Various Platforms} \label{sec:act_chek_sec}

Table~\ref{tab:rt_act_exp} shows examples of conditions for actuation commands for various cyber-physical platforms.

\subsection{Feasibility Conditions} \label{sec:timing_analysis}

Let $N_i$ be the number of actuation requests generated by task $\tau_i$ that require vetting and  $C_i^o$ be an upper bound of additional computing time required due to \ca context switching (from normal execution to secure enclave and returning the context back to normal mode) and \cb performing the checks inside the enclave. Then the execution time of $\tau_i$ can be represented as $C_i^{TEE} = C_i + N_i C_i^o$. The task $\tau_i$ is ``schedulable'' if the worst-case response time (WCRT), $R_i^{TEE}$, is less than deadline, \ie $R_i^{TEE} \leq D_i$. We can calculate an upper bound of $R_i^{TEE}$ using traditional response-time analysis~\cite{parti_see} as follows:
\begin{align} \label{eq:tz_rta}
	R_i^{TEE} = \overbrace{C_i^{TEE}}^\text{task execution time} + \hspace*{-1em} \underbrace{\sum_{\tau_h \in hp(\tau_i, \pi_p)} \hspace*{-0.5em} \left( 1+ \tfrac{D_i}{T_h} \right) C_h^{TEE}}_\text{interference from high-priority tasks}
\end{align}
where $hp(\tau_i, \pi_p) \in \Gamma_p$ denotes the set of tasks that are higher-priority than $\tau_i$ running on core $\pi_p$. The taskset $\Gamma$ is referred to as schedulable if all the tasks are schedulable, \viz $R_i^{TEE} \leq D_i, \forall \tau_i \in \Gamma$.

Let $R_i = C_i + \hspace*{-1em} \sum\limits_{\tau_h \in hp(\tau_i, \pi_p)} \hspace*{-0.5em} \left( 1+ \tfrac{D_i}{T_h} \right) C_h$ denotes the vanilla response time (\ie when there is no  checking). Notice that the task $\tau_i$ will miss its deadline if $R_i^{TEE} > D_i$. 
 From  Eq.~(\ref{eq:tz_rta}) we can deduce that $\tau_i$ will miss its deadline if the following  condition holds:
$O_i > D_i - R_i$
where $O_i = N_i C_i^o +  \hspace*{-1em} \sum\limits_{\tau_h \in hp(\tau_i, \pi_p)} \hspace*{-0.5em} \left( 1+ \tfrac{D_i}{T_h} \right) N_i C_h^o$ is the total overhead for checking the actuation commands including context switching in/out of the secure enclave.

\subsection{Linear Programming Formulation for solving the Game} \label{sec:game_linear_formulation}

We can obtain the probability distributions for selecting the elements from $X_i$ (\ie the set of all combinations of choosing $K_i$ requests from total $N_i$ actuation commands) for a given attacker strategy $l$ (that maximizes the system reward)
by forming the following linear program:
\begin{maxi!} 
	{x_i^j}{ \sum\nolimits_{j=1}^{|X_i|} x_i^j \lambda_i^{j,l}  }{}{} \label{eq:game_lp_obj}
	\addConstraint{\forall l^\prime \in [1, |Q_i|],~~}  {\sum\nolimits_{j=1}^{|X_i|} x_i^j \zeta_i^{j,l} \geq \sum\nolimits_{j=1}^{|X_i|} x_i^j \zeta_i^{j,l^\prime}} \label{eq:game_lp_con_1}
	\addConstraint{\sum\nolimits_{j=1}^{|X_i|}x_i^j} {=1} \label{eq:game_lp_con_2}
	\addConstraint{x_i^j} {> 0,~\forall j \in [1, |X_i|]} \label{eq:game_lp_con_3}
\end{maxi!}
The objective function in Eq.~(\ref{eq:game_lp_obj}) maximizes the total system reward. 
The constraint in Eq.~(\ref{eq:game_lp_con_1}) ensures that the current (\eg $l$-th) strategy results in higher cost for the attacker when compared to other adversarial strategies. The constraint in Eq.~(\ref{eq:game_lp_con_2}) ensures the sum of probability distributions equal to unity and the last constraint in Eq.~(\ref{eq:game_lp_con_3}) ensures non-zero probabilities so that all combinations of the actuation commands from $X_i$ can be selected. 

Let $[x_i^j]_{j=1:|X_i|}(l)$ denote the solution obtained from the linear programming formulation for the $l$-th adversarial strategy. Then, from all feasible strategies $l$ (where $1 \leq l \leq |Q_i|$) we choose the one (say $l^*$) that maximizes the objective value in Eq.~(\ref{eq:game_lp_obj}), \ie $l^* = \underset{1 \leq l \leq |Q_i|}{\operatorname{argmax}}  \sum\nolimits_{j=1}^{|X_i|} x_i^j \lambda_i^{j,l}$. The variables $[x_i^j]_{j=1:|X_i|}(l^*)$ obtained by solving the corresponding linear program gives us the probability distributions of selecting $K_i$ subset of commands from a total of $N_i$ commands. The game-theoretical analysis shows that the probability distributions obtained by solving the $l^*$-th linear program will be optimal for the task $\tau_i$ (\ie maximizes system reward)~\cite{conitzer2006computing,paruchuri2007efficient}.

For a given strategy $l$, the above linear programming formulation can be solved using standard off-the-shelf optimization solvers~\cite{cvxopt,python_mip} in polynomial time. Since the strategy set $Q_i$ is finite by definition, we can calculate the optimal probability distributions (\eg $[x_i^j]_{j=1:|X_i|}(l^*)$) in a finite amount of time since it is polynomial in the total number of adversarial strategies.

\subsection{Calculation of Maximum Feasible Number of Commands} \label{sec:alg_bin_search_appsec}

The pseudocode for calculating the maximum number of commands ($K_i$) that can be checked per job while guaranteeing feasibility is presented in Algorithm~\ref{alg:act_verif_log_search}. 

\renewcommand{\algorithmicforall}{\textbf{for each}}
    \renewcommand\algorithmiccomment[1]{%
 {\it /* {#1} */} %
}
\renewcommand{\algorithmicrequire}{\textbf{Input:}}
    \renewcommand{\algorithmicensure}{\textbf{Output:}}

		\begin{algorithm}[!t]
        
			\begin{algorithmic}[1]
			 \begin{footnotesize}

					\STATE Define $K_i^l := N_i^{min}, ~~ K_i^r := N_i, ~~ K_i^c := 0$
					\STATE Set $\widehat{\mathcal{K}}_i := \{N_i^{min}\}$  ~~ \COMMENT{Initialize a variable to store feasible values}
					\WHILE{$K_i^l \leq K_i^r$}
					\STATE Update $K_i^c := \lfloor \frac{K_i^l + K_i^r}{2} \rfloor$
					
					\IF{$\exists \tau_l \in lp(\tau_i, \pi_p)$ such that $\tau_l$ is \textit{not schedulable} with $K_i = K_i^c$}
					\STATE \COMMENT{Decrease verification load to make the taskset schedulable}
					\STATE Update $K_i^r := K_i^c - 1$
					\ELSE
					\STATE \COMMENT{Taskset is schedulable with $K_i^c$}
					\STATE $\widehat{\mathcal{K}}_i := \widehat{\mathcal{K}}_i \cup \{ K_i^c \}$  ~~ \COMMENT{Add $K_i^c$ to the feasible list}
					\STATE \COMMENT{Check schedulability with larger $K_i$ for next iteration}
					\STATE Update $K_i^l := K_i^c + 1$ 
					\ENDIF

					\ENDWHILE
                   	
                    
                    
                    \STATE \COMMENT{return the maximum from the set of feasible values}
                    
                    \STATE \textbf{return} $\max \big(\widehat{\mathcal{K}}_i \big)$

				\end{footnotesize}
				 
			\end{algorithmic}
			\caption{Calculation of Maximum Feasible Actuation Requests for a Given Task $\tau_i \in \Gamma_p$}
 \label{alg:act_verif_log_search}
		\end{algorithm}

\subsection{Design-time Tests for Integrating Actuation Checking in Existing Systems} \label{sec:sched_plot}

\begin{figure}[!t]
	\centering
	\includegraphics[scale=0.50]{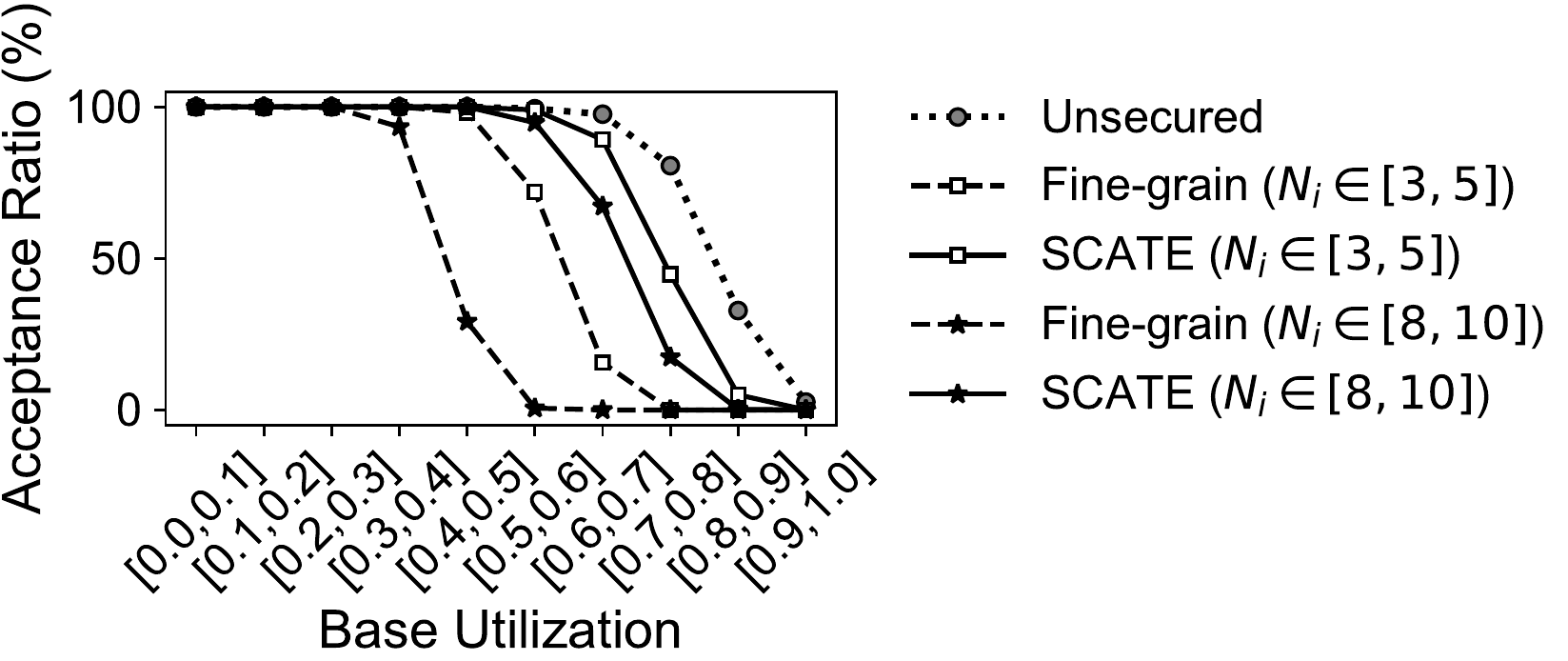}
	\caption{Impact on schedulability for the three schemes: \ci Unsecured (does not check any actuation commands), \cii Fine-grain (checks all the commands) and \cii \pnametee (selectively checks a subset of commands). We varied the number of actuation commands ($N_i$) and considered the following two scenarios: $N_i \in [3,5]$ and $N_i \in [8,10]$, $\forall \tau_i \in \Gamma$. Fine-grained checking can reduce schedulability significantly (specially if tasks have large number of actuation requests) due to increased validation overheads.} 
	\label{fig:sched}
\end{figure}

We also performed experiments to show the impact of integrating TEE-based actuation checking mechanisms (\eg \pnametee and the fine-grain scheme) in an existing system. 
For this, we use the ``schedulability'' metric introduced in Sec.~\ref{sec:syn_results}. To demonstrate the effect of schedulability for  a large number of tasksets with different parameters, let us now introduce the notion of \textit{acceptance ratio} that is defined as the number of schedulable tasksets over the total number of generated tasksets (\eg $500$ for a given utilization group in our setup as explained in Sec.~\ref{sec:syn_experiments}).

In Fig.~\ref{fig:sched} we compare the performance of difference schemes in terms of acceptance ratio (y-axis in the figure).
The x-axis shows the normalized base-utilization $\tfrac{U}{P}$ where $U = \sum_{\tau_i \in \Gamma} \tfrac{C_i}{T_i}$ (\ie taskset utilization without any actuation checking). As expected, schedulability drops for high utilization cases since less number of tasks meet their deadlines due to increased load.  While non-secure execution (\ie when there is no command verification) results in better schedulability due to reduced utilization, it \textit{does not provide any security guarantee}. The fine-grain checking, while providing better security (since it verifies every request), performs poorly in terms of meeting the timing guarantees  specifically for highly loaded systems.
 (\ie less number of tasks found to be schedulable) due to more validation overheads. In contrast, \pnametee provides better schedulability with (slight) QoS/security degradation as we demonstrate in Sec.~\ref{sec:syn_results}. The designers of the systems can use the results presented here to analyze the feasibility of integrating TEE-based checking method in their target platforms.

\subsection{Impact of Physical Inertia} \label{sec:phys_inertia}

\begin{figure}
	\centering
	\includegraphics[scale=0.480]{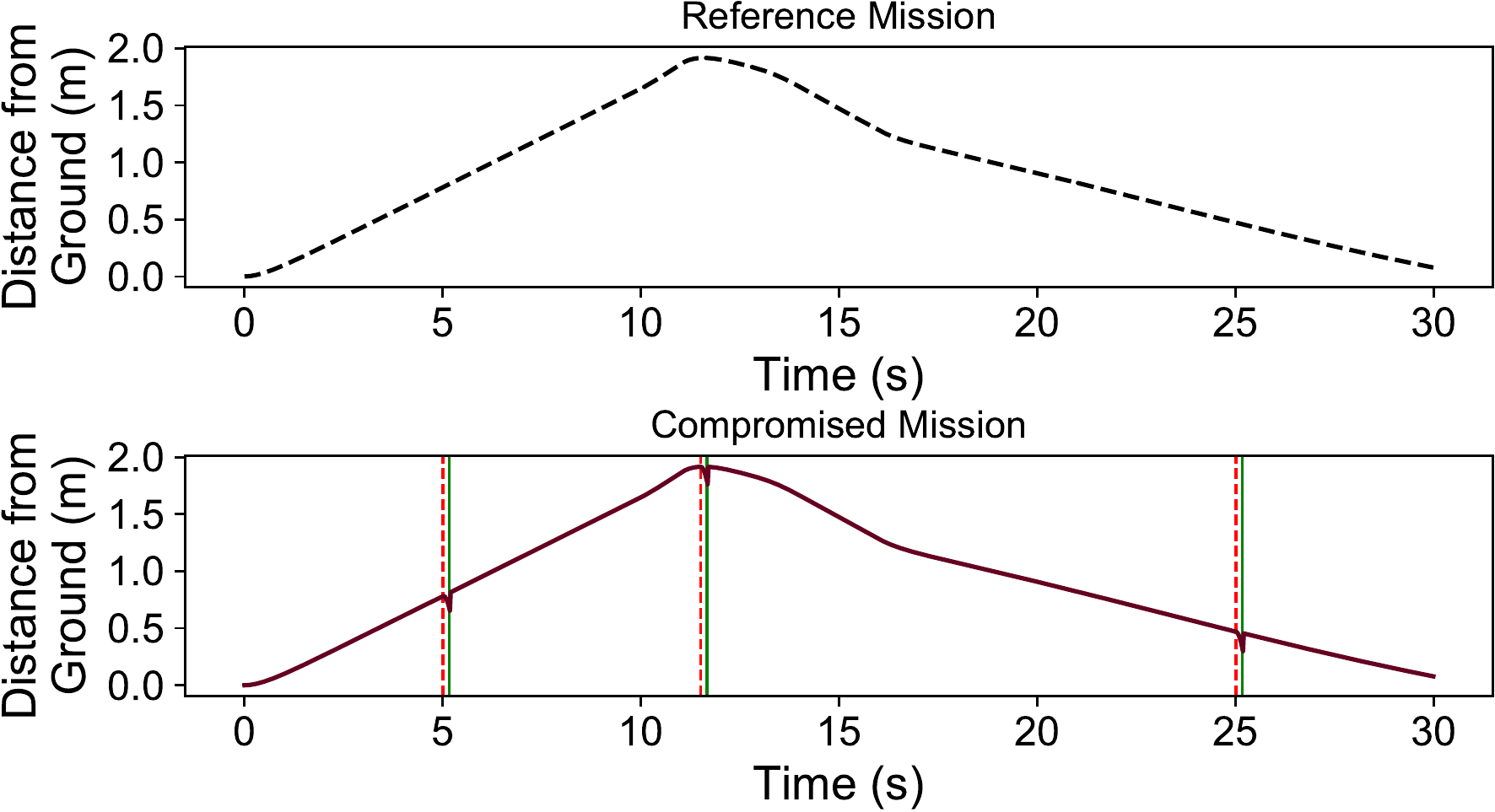}
	\caption{Effect of physical inertia in cyber-physical applications. The top plot shows the expected altitudes of the drone during mission. The bottom plot presents the altitudes during attacks. Red dashed vertical lines show the time instance when the attack is triggered and green solid vertical lines show the time instance when \pnametee detects it. While there exist slight drifts in altitudes due to delayed detection (between red and green vertical lines in the right plot), it does not jeopardize the safety (\ie the drone did not crash).} 
	\label{fig:phy_iner}
\end{figure}

We now present the impact of physical inertia for detecting attacks in \pnametee. For instance, consider a simple drone example. The baseline safety requirement for a drone is to not crash into the ground. We use existing quad-copter  models~\cite{luukkonen2011modelling} and simulate the dynamics of the drone for $30$ seconds (x-axes in Fig.~\ref{fig:phy_iner}). In this mission, the drone takes-off from the ground and then lands in the target position. The y-axes in Fig.~\ref{fig:phy_iner} represent altitudes of the drone (\ie distance from the ground) during the mission. The top plot of Fig.~\ref{fig:phy_iner} shows the corresponding altitudes over time during the normal quad-copter operation.  To demonstrate malicious activity, we injected attacks that sent false commands to turn off the propellers (bottom plot of Fig.~\ref{fig:phy_iner}). In particular, we triggered attacks at the following three points in time (red, vertical lines in the bottom side plot), \ci while the quad-copter was ascending (at $5$ sec.), \cii at the peak altitude (at $12$ sec.) and \ciii during descent (at $25$ sec.). The attacks were detected by \pnametee within $8$ task instances each (\ie $99$-th percentile values obtained from the flight controller case-study, see Table~\ref{tab:cs_res_summary}). The vertical lines in the plot represent time difference when an attack is triggered (dashed red) and when it is detected (solid green) by \pnametee. As the figure illustrates, there is a slight drift in altitude before \pnametee detects and blocks false commands. However, this delayed detection does not jeopardize safety constraints (\ie it does not drop the drone's altitude to zero) and the drone is able to complete the mission without crashing. Hence it is not inconceivable that the detection delays induced by \pnametee will be acceptable for many cyber-physical applications.

\end{document}